\documentclass[10pt,preprint]{sigplanconf}
\usepackage{graphicx}
\usepackage{subfigure}
\usepackage{url}
\usepackage{amsmath}  

\begin{document}

\authorinfo{}

\title{Exploring Linkablility of Community Reviewing}
\date{\today}\maketitle
\begin{abstract}
Large numbers of people all over the world read and contribute to various review sites. Many contributors 
are understandably concerned about privacy in general and, specifically, about linkability of their reviews 
(and accounts) across multiple review sites. In this paper, we study linkability of community-based reviewing 
and try to answer the question: {\em to what extent are "anonymous" reviews linkable, i.e., highly likely 
authored by the same contributor?}
Based on a very large set of reviews from one very popular site (Yelp), we show that a high percentage of 
ostensibly anonymous reviews can be linked with very high confidence. This is despite the fact that 
we use very simple models and equally simple features set. Our study suggests that contributors 
reliably expose their identities in reviews. This has important implications for cross-referencing 
accounts between different review sites.  Also, techniques used in our study could be 
adopted by review sites to give contributors feedback about privacy of their reviews.
\end{abstract}

\section{Introduction}
\label{sec:intro}
In recent years, popularity of various types of review and community-knowledge sites has
substantially increased. Prominent examples include Yelp, Tripadvisor, Epinions, Wikipedia, Expedia and Netflix.  
They attract multitudes of readers and contributors. While the former usually greatly outnumber the latter,
contributors can still number in hundreds of thousands for large sites, such as Yelp or Wikipedia.
For example, Yelp had more than $39$ million visitors and reached $15$ million reviews 
in late 2010 \cite{yelpstat}.  To motivate contributors to provide more (and more useful/informative) reviews, 
certain sites even offer rewards \cite{yelpeliete}. 

Some review sites are generic (e.g., Epinions) while others are domain-oriented, e.g., Tripadvisor.
Large-scale reviewing is not limited to review-oriented sites; in fact,  many retail sites encourage 
customers to review their products. e.g., Amazon and Netflix. 

With the surge in popularity of community- and peer-based reviewing, more and more people contribute 
to review sites. At the same time, there has been an increased awareness with regard to personal privacy.
Internet and Web privacy is a broad notion with numerous aspects, many of which have been explored by the 
research community. However, privacy in the context of review sites has not been adequately studied. 
Although there has been a lot of recent research related to reviewing, its focus  has been mainly on 
extracting and summarizing opinions from reviews \cite{mine-peanut, mine-summarize-rev,
thumbsup} as well as determining authenticity of reviews 
\cite{opinion-spam-analysis, opinion-spam-analysis-2, opinion-spam-analysis-3}. 

In the context of community-based reviewing, contributor privacy
has several aspects: (1) some review sites do not require accounts (i.e., allow ad hoc reviews) and
contributors might be concerned about linkability of their reviews, and
(2) many active contributors have accounts on multiple review sites and prefer these accounts not be linkable. 
The flip side of the privacy problem is faced by review sites themselves: how to address spam-reviews and
sybil-accounts?

The goal of this paper is to explore linkability of reviews by investigating how close  and related are a person's 
reviews. That is, how accurately we can link a set of anonymous reviews to their original author. 
Our study is based on over $1,000,000$ reviews and 
$\simeq~2,000$ contributors  from Yelp. The results clearly illustrate that most (up to 99\% in some cases)
reviews by relatively active/frequent contributors are highly linkable. This is despite the fact that our approach 
is based on simple models and simple feature sets. For example, using only alphabetical letter distributions, 
we can link up to 83\% of anonymous reviews. We anticipate two contributions 
of this work:  (1) extensive assessment of reviews' linkability, and (2) several models that quite accurately 
link ``anonymous'' reviews.

Our results have several implications. One of them is the ability to cross-reference contributor accounts 
between multiple  review sites. If a person regularly contributes to two review sites under different accounts,
anyone can easily link them, since most people tend to consistently maintain their traits in writing reviews. 
This is possibly quite detrimental to personal privacy. 
Another implication is the ability to correlate reviews ostensibly emanating from different accounts that are 
produced by the same author. 
Our approach can thus be very useful in detecting self-reviewing and, more
generally, review spam \cite{opinion-spam-analysis} whereby one person contributes from multiple accounts
to artificially promote or criticize products or services. 

One envisaged application of our technique is to have it  integrated into review site software. 
This way, review authors could obtain  feedback indicating the degree of linkability of their reviews. 
It would then be up to each author to adjust (or not) the writing style and other characteristics.

\noindent{\bf Organization:} Section \ref{sec:background} provides background information about 
techniques used in our experiments. The sample dataset is described in Section \ref{sec:dataset} and 
study settings are addressed in Section \ref{sec:problem-setting}. Next, our analysis methodology is 
presented in Section \ref{sec:analysis}. Section \ref{sec:disc} discusses issues stemming from 
this work and Section \ref{sec:futurework} sketches out some directions for the future. 
Then, Section \ref{sec:related} overviews related work and Section\ref{sec:conclusion} 
concludes the paper.

\section{Background}
\label{sec:background}
This section provides some background about statistical tools used
in our study. We use two well-known approaches based on:
(1) Na\"ive Bayes Model \cite{nbayes}, (2) Kullback-Leibler Divergence Metric \cite{bishopbook}. 
We briefly describe them below.

\subsection{Na\"ive Bayes Model}
\label{nb-background}
Na\"ive Bayes Model (NB) is a probabilistic model based on the eponymous  
assumption stating that all features/tokens are conditionally independent given the class. 
Given tokens:  $T_1, T_2,...,T_n$ in document $D$, we compute conditional probability 
of a document class $C$ as follows:  
%
\footnotesize
\begin{align}
&P(C|D)=P(C|T_1,T_2,...,T_n)= \frac{P(T_1,T_2,...,T_n|C)P(C)}{P(T_1,T_2,...,T_n)} \nonumber
\end{align}
\normalsize
According to the Na\"ive Bayes assumption,

\footnotesize
\begin{align}
&P(T_1,T_2,...,T_n|C) = P(T_1|C)P(T_2|C).....P(T_n|C) \nonumber
\end{align}
\normalsize

Therefore,
\footnotesize
\begin{align}
&P(C|T_1,T_2,...,T_n) = \frac{P(T_1|C)P(T_2|C).....P(T_n|C)P(C)}{P(T_1,T_2,...,T_n)} \nonumber
\end{align}
\normalsize
To use NB for classification, we return the class value with maximum probability: 
\footnotesize
\begin{align}
&Class = argmax_C P(C|D) = argmax_C P(C|T_1,T_2,...,T_n) 
\label{eq:symkl}
\end{align}
\normalsize
Since $P(T_1,T_2,...,T_n)$ is the same for all $C$ values, and assuming $P(C)$ is the same for all 
class values, the above equation is reduced to:
\footnotesize
\begin{align}
&Class = argmax_C  P(T_1|C)P(T_2|C).....P(T_n|C)\nonumber
\end{align}
\normalsize
%
%

Probabilities are estimated using the Maximum-Likelihood estimator \cite{bishopbook} as follows:
\footnotesize
\begin{align}
&P(T_i|C)=\frac{Num\ of\ Occurrences\ of\ T_i\ in\ D}{Num\ of\ Occurrences\ of\ all\ Tokens\ in\ D} \nonumber
\end{align}
\normalsize
We smooth the probabilities with Laplace smoothing \cite{mlearn1} as follows:
\footnotesize
\begin{align}
&P(T_i|C)= \nonumber\\
&\frac{Num\ of\ Occurrences\ of\ T_i\ in\ D\ +\ 1}{Num\ of\ Occurrences\ of\ all\ Tokens\ in\ D\ +
\ Num\ of\ Possible\ Token} \nonumber
\end{align}
\normalsize

\subsection{Kullback-Leibler Divergence Metric}
\label{kl-background}
Kullback-Leibler Divergence (KLD) metric measures the distance between two distributions.
For any two distributions $P$ and $Q$, it is defined as:
\footnotesize
\begin{align}
D_{kl}(P\|Q)=\sum_iP(i)log(\frac{P(i)}{Q(i)}) \nonumber
\end{align}
\normalsize

KLD is always positive: the closer to zero, the closer $Q$ is to $P$. It is an asymmetrical metric,
i.e., $D_{kl}(P\|Q) \neq\ D_{kl}(Q\|P)$. To transform it into a symmetrical metric, we use the following formula (that
has been used in \cite{mal-domains-detect}):
\footnotesize
\begin{equation}
SymD_{kl}(P,Q)= 0.5 \times (D_{kl}(P\|Q) + D_{kl}(Q\|P))
\label{eq:symkl}
\end{equation}
\normalsize
Basically, $SymD_{kl}$ is a symmetrical version of $D_{kl}$ that measures the distance between two
distributions. As discussed below, it is used heavily in our study. In the rest of the paper, the term
"KLD" stands for $SymD_{kl}$\footnote{Note that, under certain conditions, 
NB and asymmetrical KLD models could be equivalent. That is,  
$argmax_{Class} P(Class|T_1,T_2,...,T_n)$ is equivalent to $argmin_{Class} D_{kl}(Token\_distribution\|Class\_distribution)$,
where $T_1,T_2,...T_n$ are the tokens of a document $D$  and $Token\_distribution$ is their derived distribution. 
The proof for this equivalency is  in \cite{mal-domains-detect}. However, this equivalence does not 
hold when we use the symmetrical version $SymD_{kl}$.}.

\section{Data Set}
\label{sec:dataset}
\begin{figure}[t]
  \centering
  \subfigure[]{\label{fig:dataset-cdf}\includegraphics[scale=0.3]{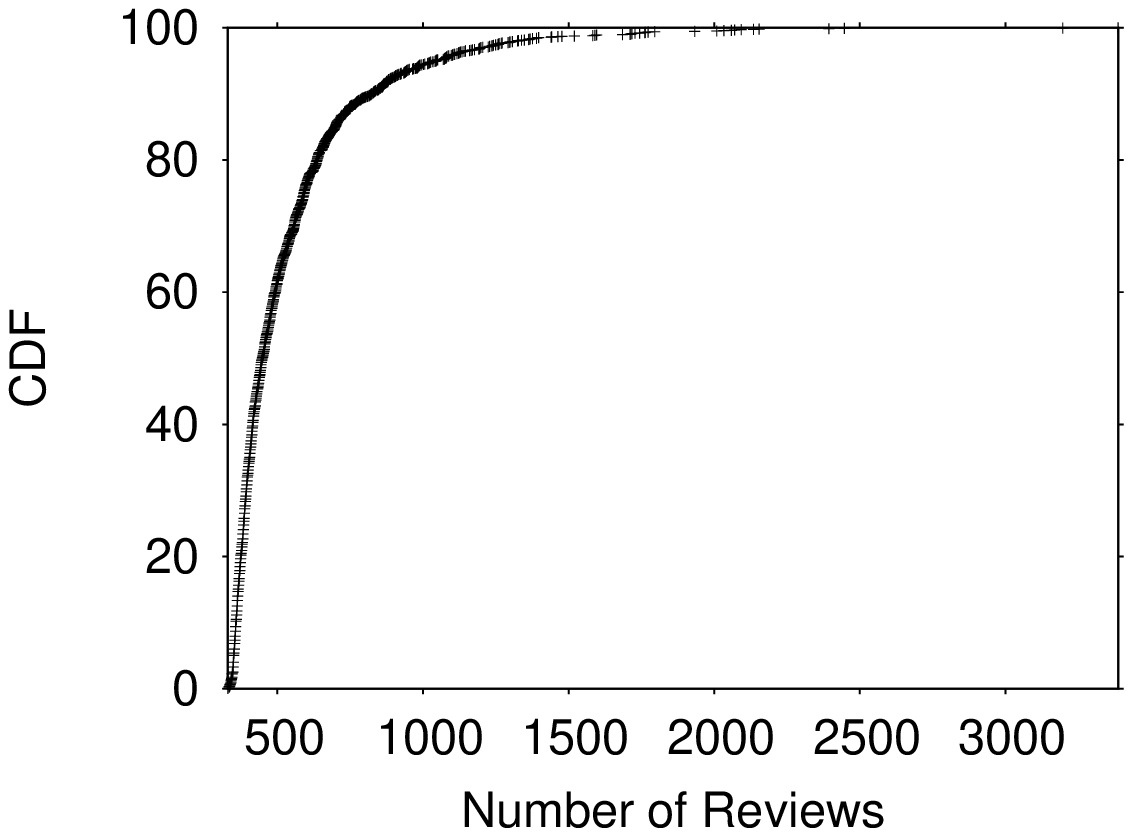}}
  \subfigure[]{\label{fig:avg-rev-size-cdf}\includegraphics[scale=0.3]{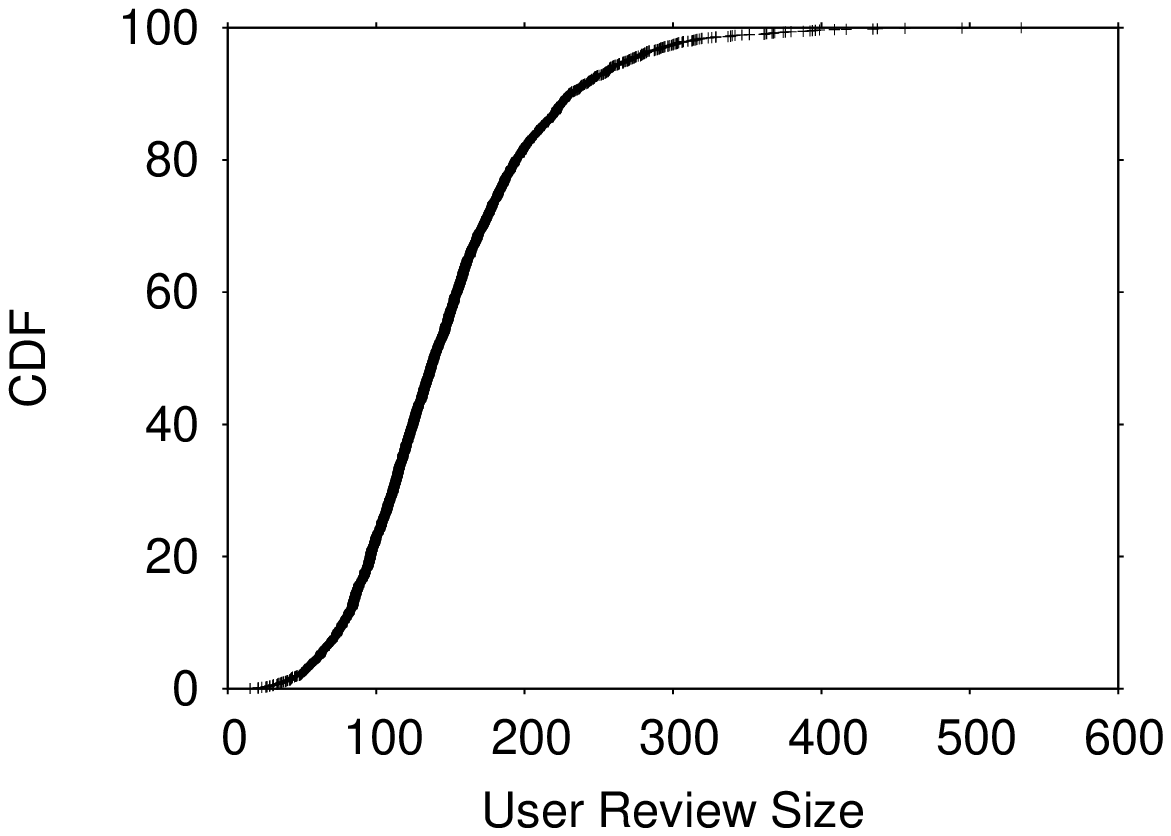}}
   \caption{CDF for: (a) number of reviews per contributor, and (b) average review size (number of words) per 
   contributor.}
\end{figure}

%
Clearly, a very large set of reviews authored by a large number of contributors is necessary
in order to perform a meaningful study. To this end, we collected $1,076,850$ reviews for $1,997$ 
contributors from \url{yelp.com}, a very popular site with many prolific contributors. As shown in Figure 
\ref{fig:dataset-cdf}, the minimum number of reviews per contributor is $330$, the maximum -- $3,387$ 
and the average -- $539$ reviews, with a standard deviation of $354$. For the purpose of this study, 
we limited authorship to prolific contributors, since this provides more useful information for the purpose
of review linkage. 

Figure \ref{fig:dataset-cdf} shows the Cumulative Distribution Function (CDF) of the 
number of reviews per contributor. 
50\% of the contributors  authored fewer than $500$ reviews and 76\% authored fewer than $600$. 
Only 6\% of the contributors exceed $1,000$ reviews. 

Figure \ref{fig:avg-rev-size-cdf} shows the CDF for average review size (number of words) per contributor. 
It shows that 50\% of the contributors write reviews shorter than $140$ words (on average) and 75\% --
have average review size smaller than $185$. Also, 97\% of contributors write reviews shorter than 
$300$ words. The  overall average review size is relatively small -- $149$ words. 

\section{Study Setting}
\label{sec:problem-setting}
As mentioned earlier, our central goal is to study linkability of relatively prolific reviewers. 
Specifically, we want to understand -- for a given prolific author -- to what extent some of his/her 
reviews relate to, or resemble, others. To achieve that, we first randomly order the reviews of each 
contributor. Then, for each contributor $U$ with $N_U$ reviews, we split the randomly ordered 
reviews into two sets: 
\begin{enumerate}
\item First $N_U - X$ reviews:  We refer to this as the \textbf{identified record} (IR) of $U$.
\item Last $X$ reviews: These reviews represent the full set of anonymous reviews of $U$ from which we derive 
several subsets of various sizes. We refer to each of these subset as an \textbf{anonymous record} (AR) 
of $U$. An AR  of size $i$ consists of the first $i$ reviews of the full set of anonymous reviews of $U$. 
We vary the AR size for the purpose of studying the user reviews linkability under different numbers of 
anonymous reviews.    
\end{enumerate}
Since we want to restrict the AR size to a small portion of the complete user reviews set, we 
restrict $X$ to 60 as this represents less than 20\% of the minimum number of reviews 
for authors in our set (330 total). We use the \textbf{identified records} (IRs) of all
contributors as the training set upon which we build models for linking anonymous reviews.
(Note that the IR size is not the same for all contributors, while the AR size is uniform.)
Thus, our problem is reduced to matching an anonymous record to its corresponding IR. 
Specifically, one anonymous record serves as input to a matching/linking model and the output is a 
sorted list of all possible account-ids (i.e., IR sets) listed in a descending order of probability, i.e., 
the top-ranked account-id
corresponds to the contributor whose IR represents the most probable match for the input
anonymous record. Then, if the correct account-id of the actual author is among top $T$ entries, 
the matching/linking model has a hit; otherwise, it is a miss.
Consequently, our study boils down to exploring matching/linking models that maximize the hit ratio 
of the anonymous records for varying values of both $T$ and AR sizes. 
We consider three values of $T$: 1 (perfect hit), 10 (near-hit) and 50 (near-miss). 
Whereas, for the AR size, we experiment with a wider range of values:   1, 5, 10, 20, 30, 40, 50 and 60. 

Even though our focus is on  the linkability of prolific users, we also attempt to assess performance of 
our models for non-prolific users. For that, we slightly change the problem setting by making 
the IR size smaller; this is discussed in Section \ref{sec:train-min-max}.
%
%

\section{Analysis}
\label{sec:analysis}

As mentioned in Section \ref{sec:background}, we use Na\"ive Bayes (NB) and
Kullback-Leibler Divergence (KLD) models.
Before analyzing the collected data, we tokenize all reviews and extract four types of tokens:
\begin{enumerate}
\item \textbf{Unigrams:} set of all single letters. We discard all non-alphabetical characters.
\item \textbf{Digrams:} set of all consecutive letter-pairs. We discard all non-alphabetical characters.
\item \textbf{Rating:} rating associated with the review. (In Yelp, this ranges between 1 and 5).
\item \textbf{Category:} category associated with the place/service being reviewed. 
There are 28 categories in our dataset,
\end{enumerate}
%
%
\noindent{\bf Why such simple tokens?} Our choice of these four primitive token types might
seem trivial or even na\"ive. In fact, initial goals of this study included more ``sophisticated''
types of tokens, such as: (1) distribution of word usage, (2) sentence length in words, and (3) punctuation
usage. We originally planned to use unigrams and digrams as a baseline, imagining that (as long as all
reviews are written in the same language -- English, in our case) single and double-letter
distributions would remain more-or-less constant across contributors. However, as our results
clearly indicate, our hypothesis was wrong.

In the rest of this section, we analyze results produced by NB and KLD models.
Before proceeding, we re-cap abbreviations and notation in Table \ref{tab:notation}.
\begin{table}[t!]
\footnotesize
\begin{center}
\begin{tabular}{|r|l|}
\hline
NB  & Na\"ive Bayes Model
\\ \hline
KLD  & Symmetrical Kullback-Leibler Divergence Model
\\  \hline
R  & Token Type: rating, unigram or digram
\\ \hline
LR & Linkability Ratio
\\ \hline
AR & Anonymous Record
\\ \hline
IR  & Identified Record (corresponding to a certain reviewer)
\\ \hline
$SymD_{KLD}(IR,AR)$  & symmetric KLD distance between $IR$ and $AR$
\\ \hline
$SymD_{KLD\_r}$  & symmetric KLD of rating tokens
\\ \hline
$SymD_{KLD\_c}$  & symmetric KLD of category tokens
\\ \hline
$SymD_{KLD\_l}$  & symmetric KLD of lexical(unigram or digram) tokens
\\ \hline
$SymD_{KLD\_r\_c}$  & symmetric KLD of rating and category tokens
\\ \hline
$SymD_{KLD\_l\_r\_c}$  & symmetric KLD of lexical, rating and category tokens
\\ \hline
\end{tabular}
\end{center}
\caption{Notation and abbreviations.}
\label{tab:notation}
\end{table}%

\subsection{Methodology}
\label{method}
We begin with the brief description of the methodology for the two models.

\subsubsection{Na\"ive Bayes (NB) Model}
\label{sec:nb-method}
For each account $IR$, we built an NB model, $P(token_i|IR)$,  from its identified record. 
Probabilities are computed using the Maximum-Likelihood estimator \cite{bishopbook} and Laplace 
smoothing \cite{mlearn1} as shown in \ref{sec:background}. We then construct four models 
corresponding to the four aforementioned token types.
That is, for each $IR$, we have $P_{unigram}$, $P_{digram}$, $P_{category}$ and $P_{rating}$.

To link an anonymous record $AR$ to an account $IR$ with respect to token type $R$, we first extract
all $R$-type tokens from $AR$, $T_{R_1}, T_{R_2}, ....T_{R_n}$ (Where $T_{R_i}$ is the $i$-th
$R$ token in $AR$).
Then, for each $IR$, we compute the probability $P_R(IR|T_{R_1}, T_{R_2}, ....T_{R_n})$.
Finally, we return a list of accounts sorted in decreasing order of probabilities. The top entry represents
the most probable match.

\subsubsection{Kullback-Leibler Divergence (KLD) Model}
\label{sec:kl-method}
We use symmetric KLD (see Section \ref{sec:background}) to compute the distance between
anonymous and identified records. To do so, we first compute distributions of all records, as follows:
\footnotesize
\begin{align} 
&Dist\_token(Token_i) = \frac{Num\ of\ Occurrences\ of\ Token_i}{Num\ of\ Occurrences\ of\ all\ Tokens}\nonumber
\end{align}
\normalsize
To avoid division by 0, we smooth distributions via Laplace smoothing \cite{mlearn1}, as follows:
\footnotesize
\begin{align}
&Dist\_token(Token_i) = \nonumber \\
&\frac{Num\ of\ Occurrences\ of\ Token_i\ +\ 1}{Num\ of\ Occurrences\ of\ all\ Tokens+Num\ of\ Possible\ Tokens}\nonumber
\end{align}
\normalsize
As before, we compute four distributions.
To link $AR$ with respect to token type $R$, we compute $SymD_{kl}$ between the distribution of
$R$ for  $AR$ and the distribution of $R$ for each $IR$. Then, we return a list sorted in
ascending order of $SymD_{KLD}(IR,AR)$ values. The first entry represents the account with the most likely
match.

\subsection{Study Results}
\label{sec:results}
We now present the results corresponding to our four  tokens.
Then, in the next section, we experiment with some combinations thereof.

\subsubsection{Non-Lexical: Rating and Category}
\label{sec:rating-results}
\begin{figure}[t]
  \centering
  \subfigure[]{\label{fig:rate-nb-1-60}\includegraphics[scale=0.3]{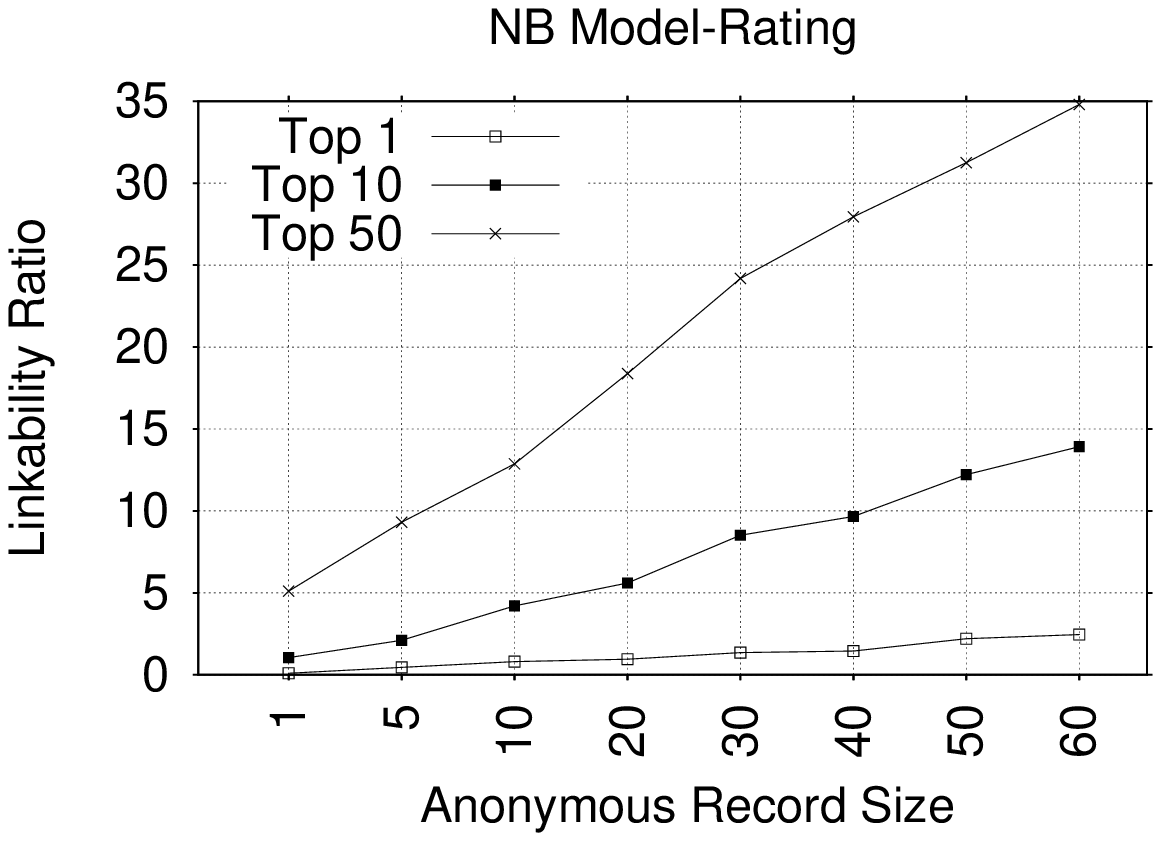}}
  \subfigure[]{\label{fig:rate-kl-1-60}\includegraphics[scale=0.3]{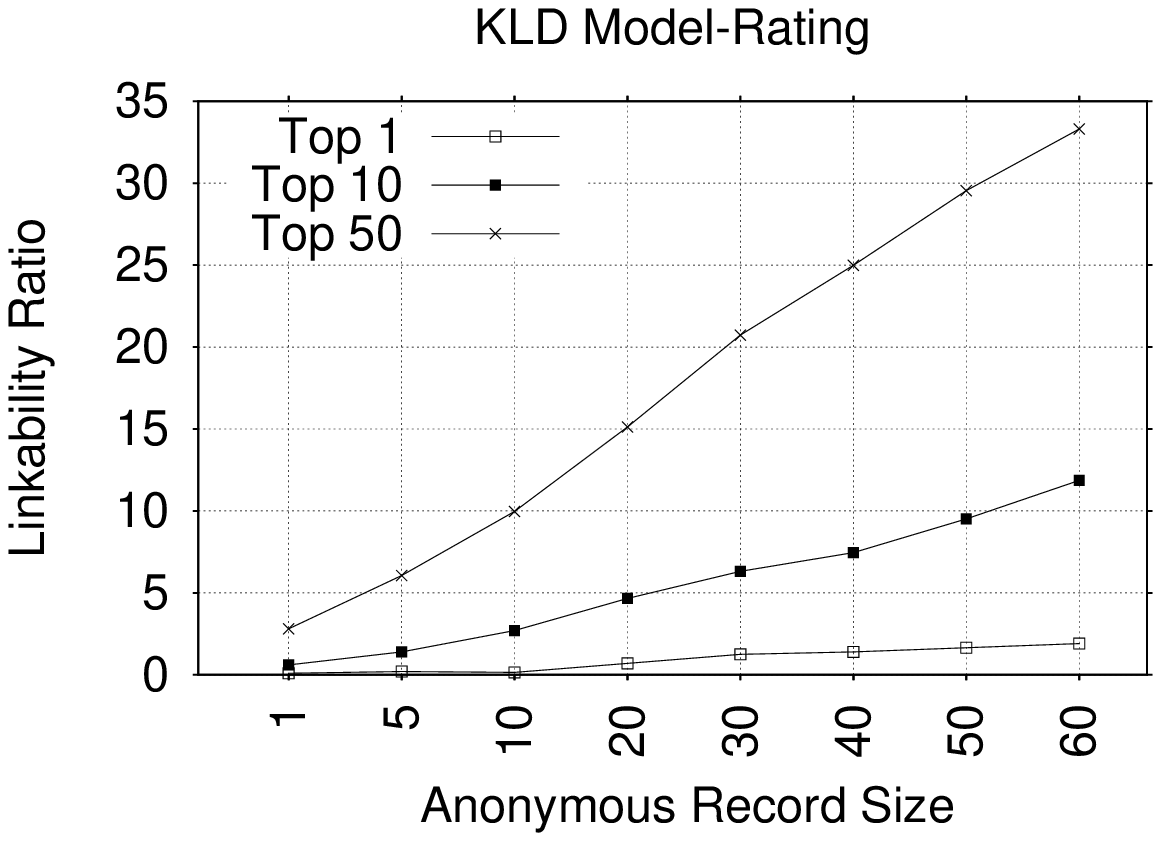}}
  \subfigure[]{\label{fig:cat-nb-1-60}\includegraphics[scale=0.3]{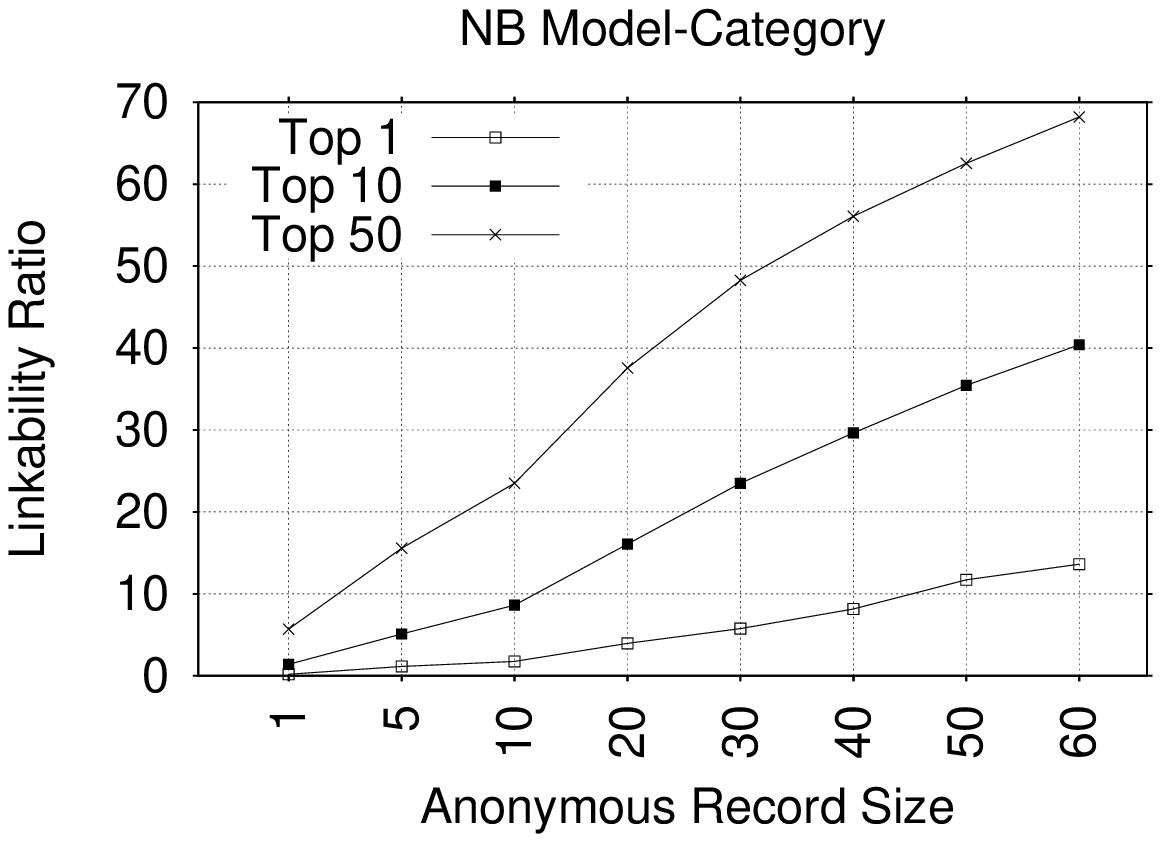}}
  \subfigure[]{\label{fig:cat-kl-1-60}\includegraphics[scale=0.3]{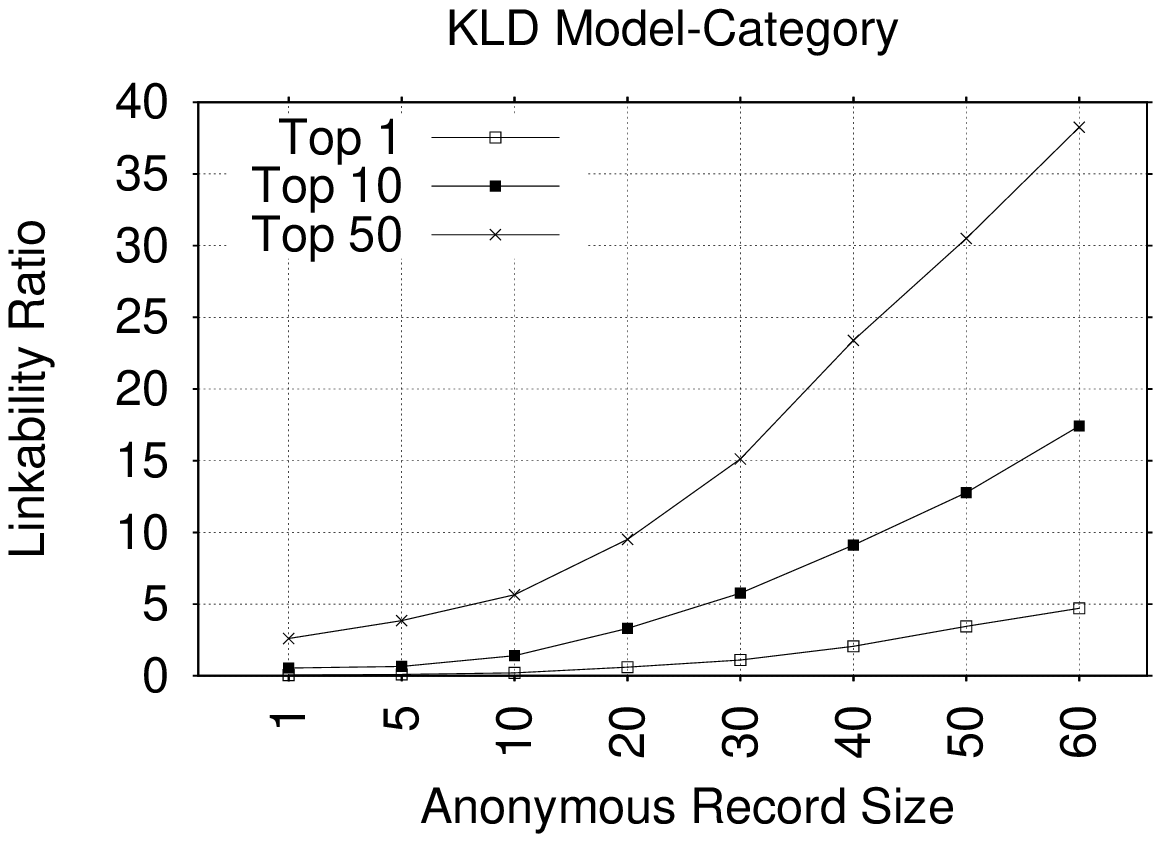}}
  \caption{LRs for NB and KLD models rating and category tokens}
  \label{fig:rate-cat-1-60}
\end{figure}

Figure \ref{fig:rate-cat-1-60}  shows Top-1, Top-10, and Top-50 plots of 
the linkability ratios (LRs) for NB and KLD models 
for several anonymous record sizes when either rating or category is used as the token. 
Not surprisingly, an increase in the anonymous record size causes an increase in the LR. 
Figures \ref{fig:rate-nb-1-60} and \ref{fig:rate-kl-1-60} show LRs when rating token alone is used. 
In the Top-1 plot, LRs are low and the highest ratio is 2.5\%/1.9\% in NB/KLD for an anonymous record size of 60. 
However, in Top-10 and Top-50 plots, LRs become higher and reach 13.9\%(11.9\%) and 34.8\%(33.3\%) in 
Top-10 and Top-50 plots, respectively, in NB(KLD) for the same anonymous record size.
Figures \ref{fig:cat-nb-1-60} and \ref{fig:cat-kl-1-60} show LRs for the category token. 
In Top-1, the highest LR is  13.6\%/4.7\% in NB/KLD for anonymous record size of 60. 
A significant increase occurs in LRs in Top-10 and Top-50 plots: 40.4\%(17.4\%) and 68.2\%(38.3\%),
respectively, in NB(KLD) model. The category is clearly more effective than the rating token. 
Additionally, we observe that NB performs better than KLD model, especially, for the category token. 

We conclude that rating- and category-based models are only somewhat helpful, yet
insufficient to link accounts for many anonymous records. However, it turns out that 
they are quite useful when combined with other lexical tokens, as discussed in 
Section \ref{sec:combine-model} below.

\subsubsection{Lexical: Unigram and Digram}
\label{sec:uni-di-results}
\begin{figure}[t]
  \centering
  \subfigure[]{\label{fig:ng-1-nb-1-60}\includegraphics[scale=0.3]{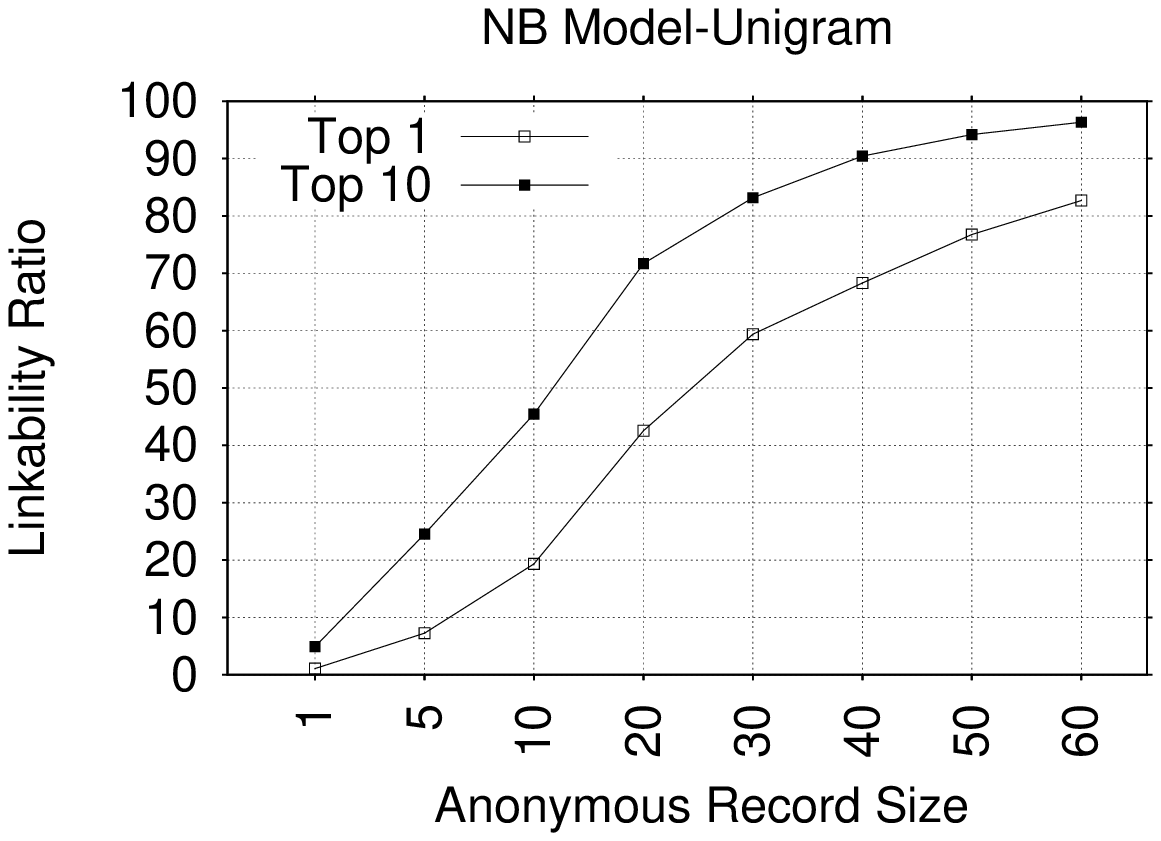}}
  \subfigure[]{\label{fig:ng-1-kl-1-60}\includegraphics[scale=0.3]{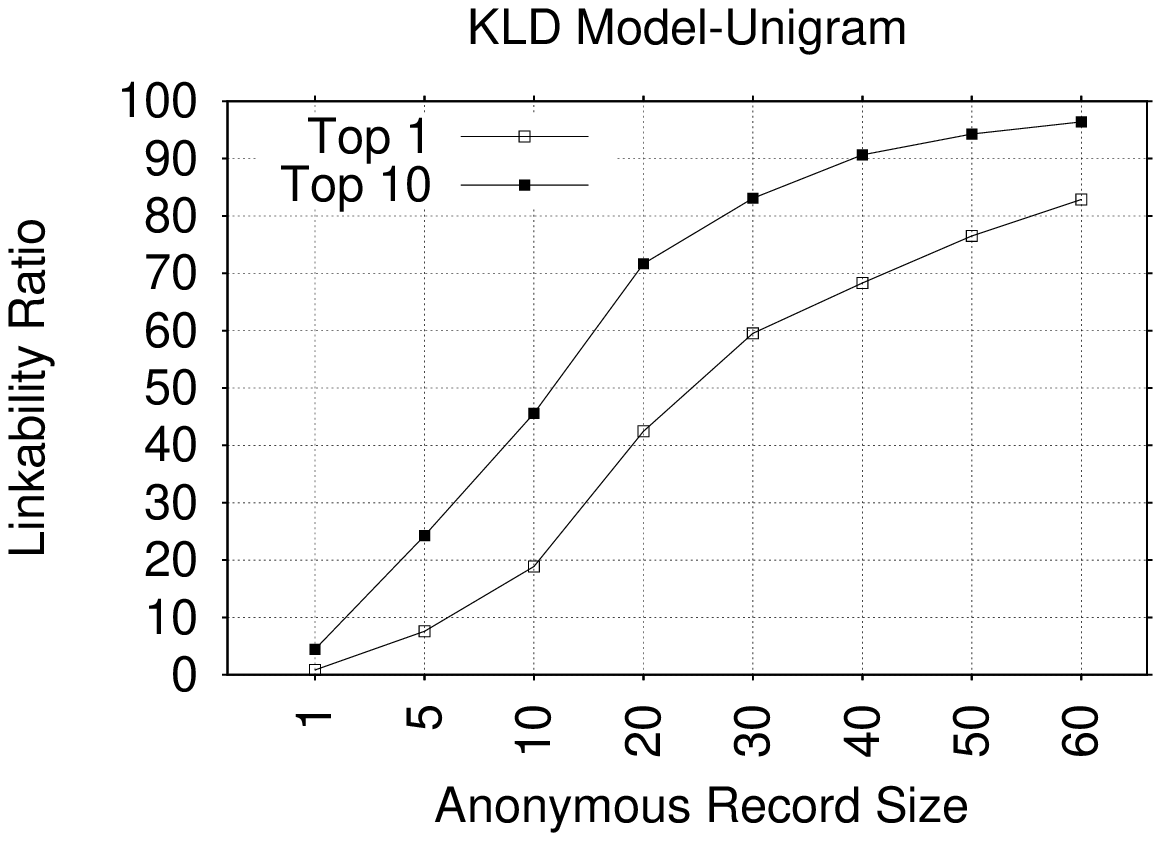}}
  \subfigure[]{\label{fig:ng-2-nb-1-60}\includegraphics[scale=0.3]{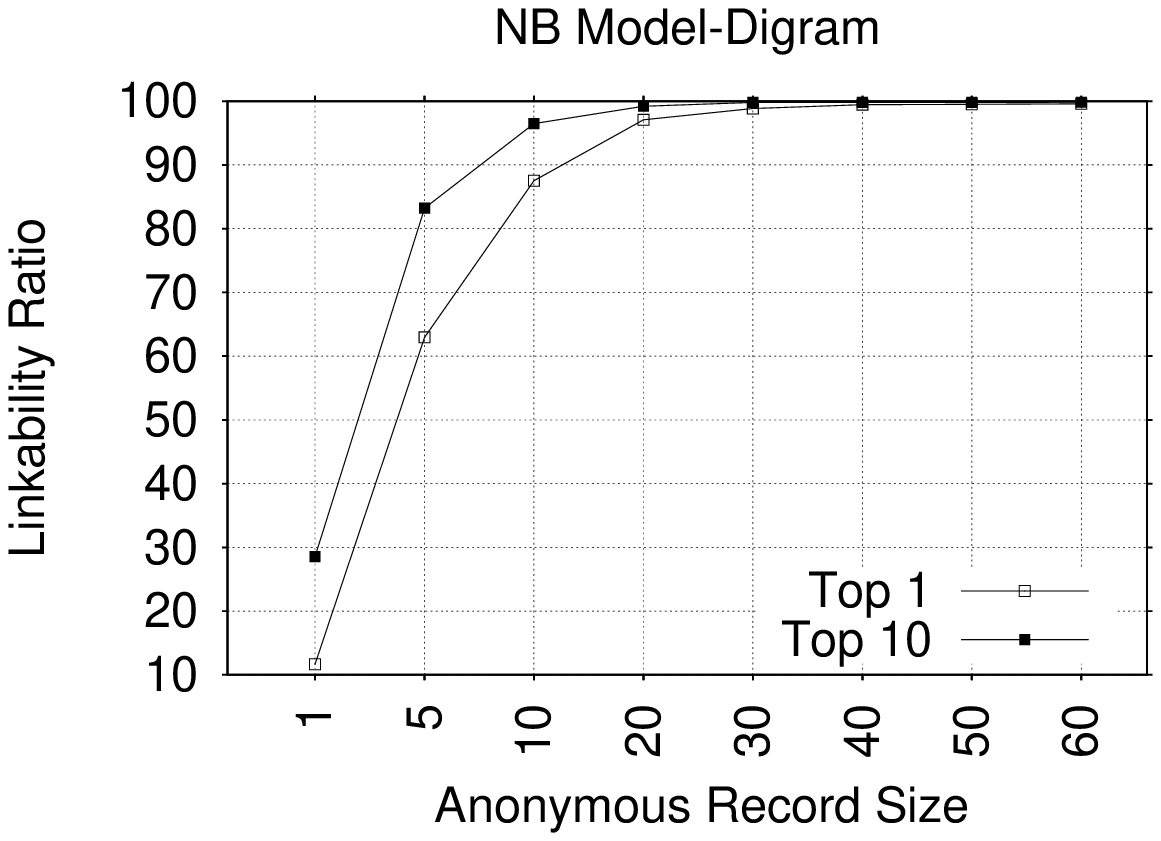}}
  \subfigure[]{\label{fig:ng-2-kl-1-60}\includegraphics[scale=0.3]{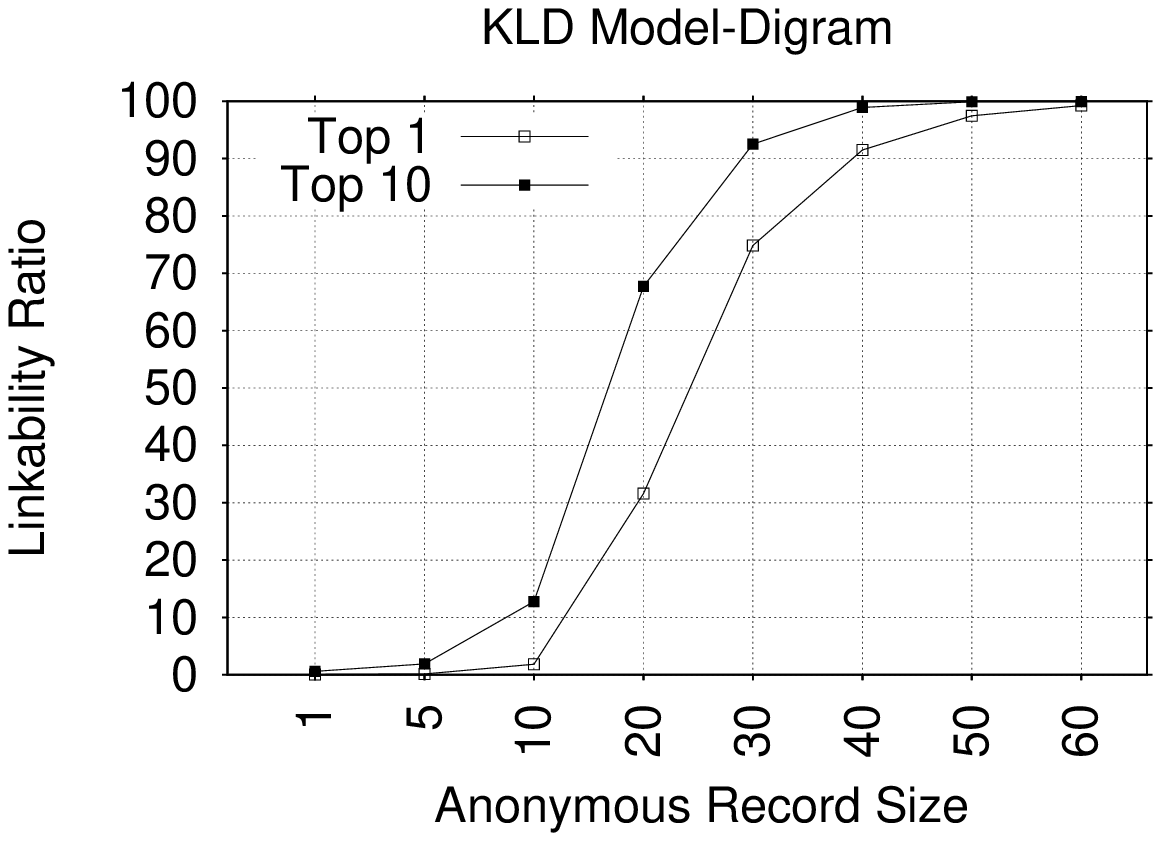}}
  \label{fig:uni-di-1-60}  
  \caption{LRs of NB and KLD models for unigrams and digrams}
\end{figure}

Figures \ref{fig:ng-1-nb-1-60} and \ref{fig:ng-1-kl-1-60}
depict LRs (Top-1 and Top-10) for NB and KLD with the unigram token.
As expected, with the increase in the anonymous record size, the LR grows: it is high in both Top-1 and Top-10 plots.
For example, in Top-1 of both figures, the LRs are around: 19\%, 59\% and 83\% for anonymous record sizes 
of 10, 30 and 60, respectively.  Whereas, in Top-10 of both figures, the LRs are around: 45.5\%, 83\% and  
96\% for same record sizes. This suggests that reviews are highly linkable based on trivial 
single-letter distributions. Note that the two models exhibit similar performance.

Figures \ref{fig:ng-2-nb-1-60} and \ref{fig:ng-2-kl-1-60} consider the digram token.
In both models, the LR is impressively high: it gets as high as 99.6\%/99.2\% in Top-1 for NB/KLD 
for an AR size of 60. For example, the Top-1 LRs in NB are: 11.7\%, 62.9\%, 87.5\% and 97.1\%, for respective
AR sizes of 1, 5, 10 and 20. Whereas, in KLD, the Top-1 LRs for record sizes of 10, 30 and 60 are: 
1.9\%,74.9\% and 99.2\%, respectively. 

Unlike unigrams -- where LRs in both models are comparable -- KLD in digram
starts with LRs considerably lower than those of  NB. However, the situation 
changes when the record size reaches 50,
with KLD performing comparable to NB. One reason for that could be that KLD improves
when the distribution of ARs is more similar to that of corresponding identified records; this
usually occurs for large record sizes, as there are more tokens.

Not surprisingly, in both lexical and non-lexical models, larger AR sizes entail higher LRs. 
With NB, a larger record size implies that, a given AR has more tokens in 
common with the corresponding IR. Thus, an increase in the prediction probability $P(IR|T_1,T_2,...T_n)$. 
For KLD, a larger record size causes the distribution derived from the AR to be more similar to the 
one derived from the corresponding IR.

\subsection{Improvement I: Combining Lexical with non-lexical Tokens}
\label{sec:combine-model}
In an attempt to improve the LR, we now combine the non-lexical token with its lexical counterparts.

\subsubsection{Combining Tokens Methodology}
\label{sec:method-kl-combine}
This is straightforward in the NB. We simply increase the list of tokens in the unigram- or digram-based NB
by adding the non-lexical tokens. Thus, for every AR, we have
$P(lexical\_token_i|IR)$, $P(category\_token_i|IR)$ and $P(rate\_token_i|IR)$.

Combining non-lexical with lexical tokens in KLD is less clear.
One way is to simply average $SymD_{KLD}$ values for both token types. However, this might
degrade the performance, since lexical distributions convey much more information than their
non-lexical counterparts. Thus, giving them the same weight would not yield better results. 
Instead, we combine them using a weighted average. First, we compute the weighted average 
of rating and category $SymD_{KLD}$:
\footnotesize
\begin{align} 
&SymD_{KLD\_r\_c}(P,Q) = \nonumber \\
&\beta \times SymD_{KLD\_r}(P,Q) + (1 - \beta) \times SymD_{KLD\_c}(P,Q)\nonumber 
\end{align}
\normalsize
Then, we combine the above with $SymD_{KLD}$ of the lexical tokens to compute 
the final weighted average:
\footnotesize
\begin{align} 
&SymD_{KLD\_l\_r\_c}(P,Q) = \nonumber\\
&\alpha \times SymD_{KLD\_l}(P,Q) + (1 - \alpha) \times SymD_{KLD\_r\_c}(P,Q)\nonumber 
\end{align}
\normalsize
Thus, our goal is to get the right $\beta$ and $\alpha$ values. Intuitively, $SymD_{kl\_lexical}$ 
should have more weight as it carries more information. Since there is no clear
way of assigning weight values, we experiment with several choices and pick the one with the 
best performance; we discuss the selection process below. We experiment only within the IR 
set and then verify the results generalizes to the AR. This is done as follows:

First, for every IR, we allocate the last 30 reviews as a testing record and the remainder -- 
as a training record. Then, we experiment with $SymD_{KLD\_r\_c}$ using several $\beta$ values and 
set $\beta$ to the value that yields the highest LR based on the tested records. Then, 
we experiment with $SymD_{KLD\_l\_r\_c}$ using several $\alpha$ values and, similarly, 
pick the one with the highest LR.
\begin{figure}[t]
  \centering
  \subfigure[]{\label{fig:train-r-c-subset}\includegraphics[scale=0.3]{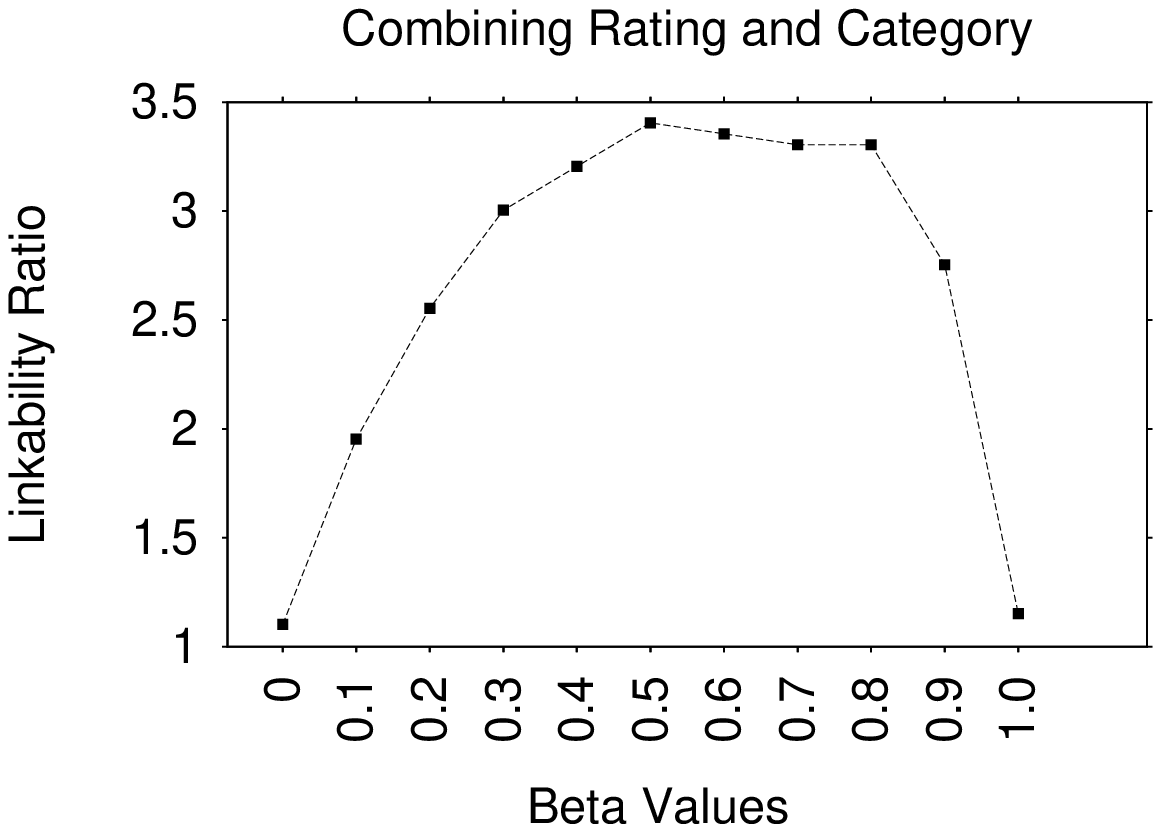}}
  \subfigure[]{\label{fig:train-1-subset}\includegraphics[scale=0.3]{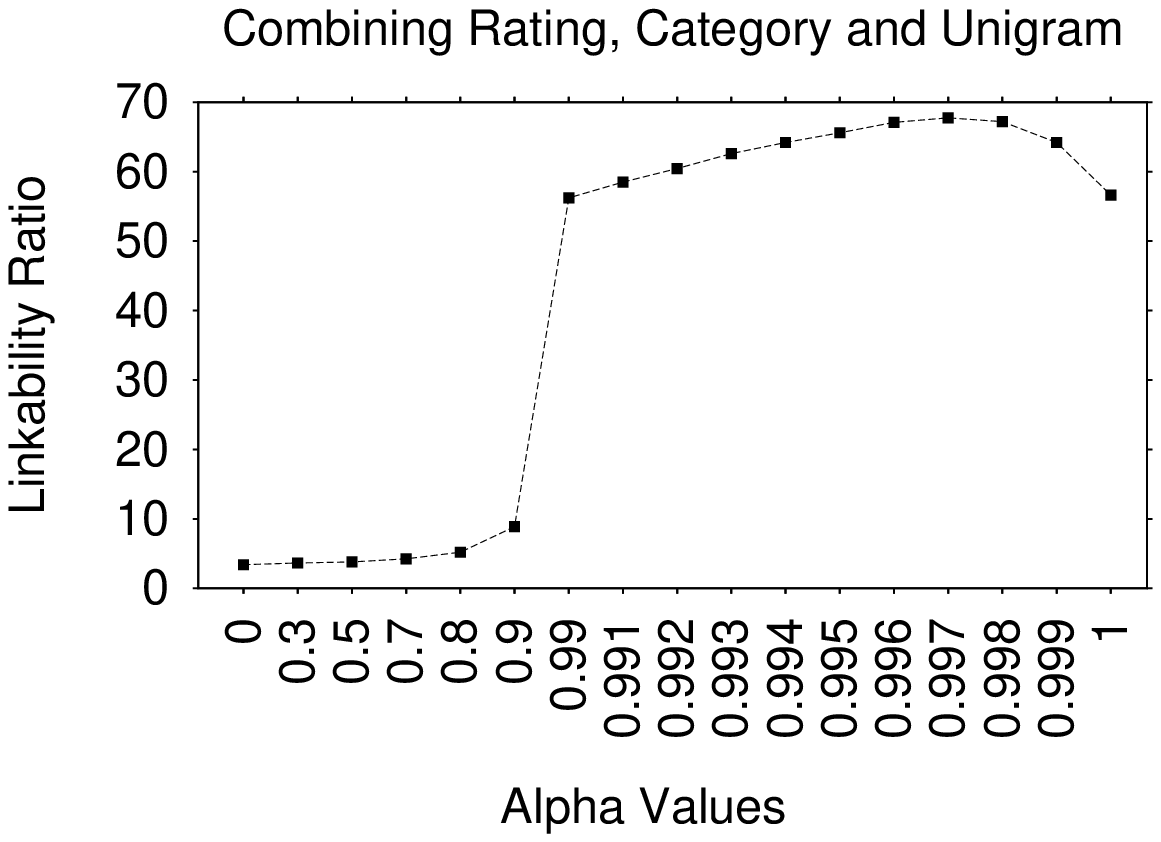}}
  \subfigure[]{\label{fig:train-2-subset}\includegraphics[scale=0.3]{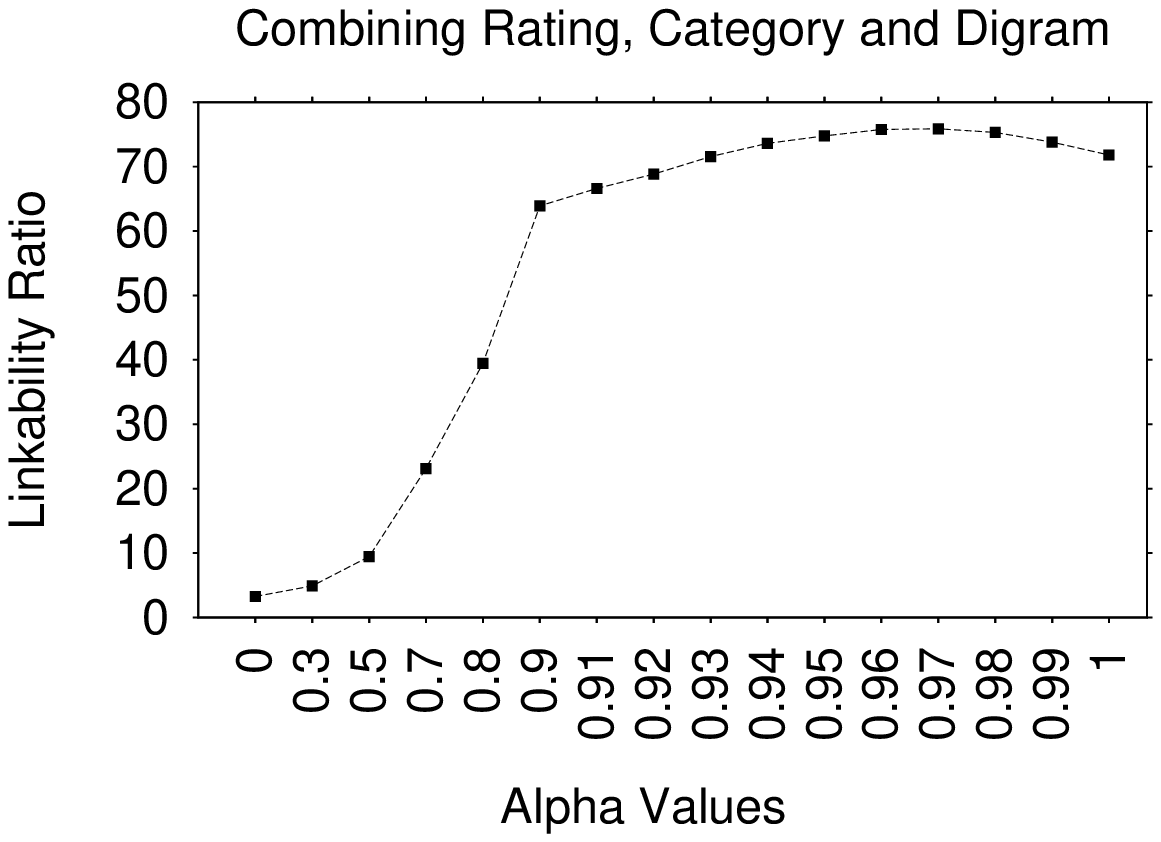}}
  \caption{Results of combining different tokens using different $\beta$ and $\alpha$ values}
  \label{fig:train-subset}
\end{figure}

Since $\beta$ or $\alpha$ could assume any values, we need to restrict their choices. For $\beta$, 
we postulate that its optimal value is close to $0.5$ since LRs for rating and category are comparable. 
Thus, we experiment with a range of values, from $0$ to $1.00$ in $0.1$ increments. 
For $\alpha$, we expect the optimal value to exceed $0.9$, since the LR for lexical tokens is significantly 
higher than for non-lexical ones. Therefore, we experiment with the weighted average by varying
$\alpha$ between $0.9$ and $0.99$ in $0.01$ increments. 

If the values exhibit an increasing trend (i.e., $SymD_{KLD\_l\_r\_c}$ at $\alpha$ of 0.99 is the 
largest in this range) we continue experimenting in the $0.99--1.00$ range in $0.001$ increments.
Otherwise, we stop. For further verification, we also experiment with smaller $\alpha$ values: 
$0.0, 0.3, 0.5, 0.7, \mbox{and}\ 0.8$, all of which yield LRs significantly lower than 0.9 
for both the unigram and digram. We acknowledge that we may be missing $\alpha$ or $\beta$ 
values that could further optimize $SymD_{KLD\_l\_r\_c}$. However, results in sections 
\ref{sec-results-rate-cate-comb} and \ref{sec-results-uni-di-rate-cate-comb}, show that our selection yields good results.  

Figure \ref{fig:train-r-c-subset} shows LRs (Top-1) for $\beta$ values. The LR gradually increases until it tops off at $3.4\%$ with $\beta=0.5$ and then it gradually decreases. Figure \ref{fig:train-1-subset} shows LRs (Top-1) for $\alpha$ values in the unigram case. The LR has an increasing trend until it reaches $67.8\%$ with $\alpha=0.997$ and then it decreases. Figure \ref{fig:train-2-subset} shows LRs (Top-1) for $\alpha$ values in the diagram case where it tops off at $75.9\%$ with $\alpha=0.97$. Thus, the final values are $0.5$ for $\beta$ and $0.997$/$0.97$ for $alpha$ in the unigram/digram case. Even though we extract $\alpha$ and $\beta$ values by testing on a record size of 30, the results in following sections show that the derived weights are effective when tested on ARs of other sizes.  

\subsubsection{Combining Rating and Category - Results}
\label{sec-results-rate-cate-comb}

\begin{figure}[t]
  \centering
  \subfigure[]{\label{fig:r-c-nb}\includegraphics[scale=0.3]{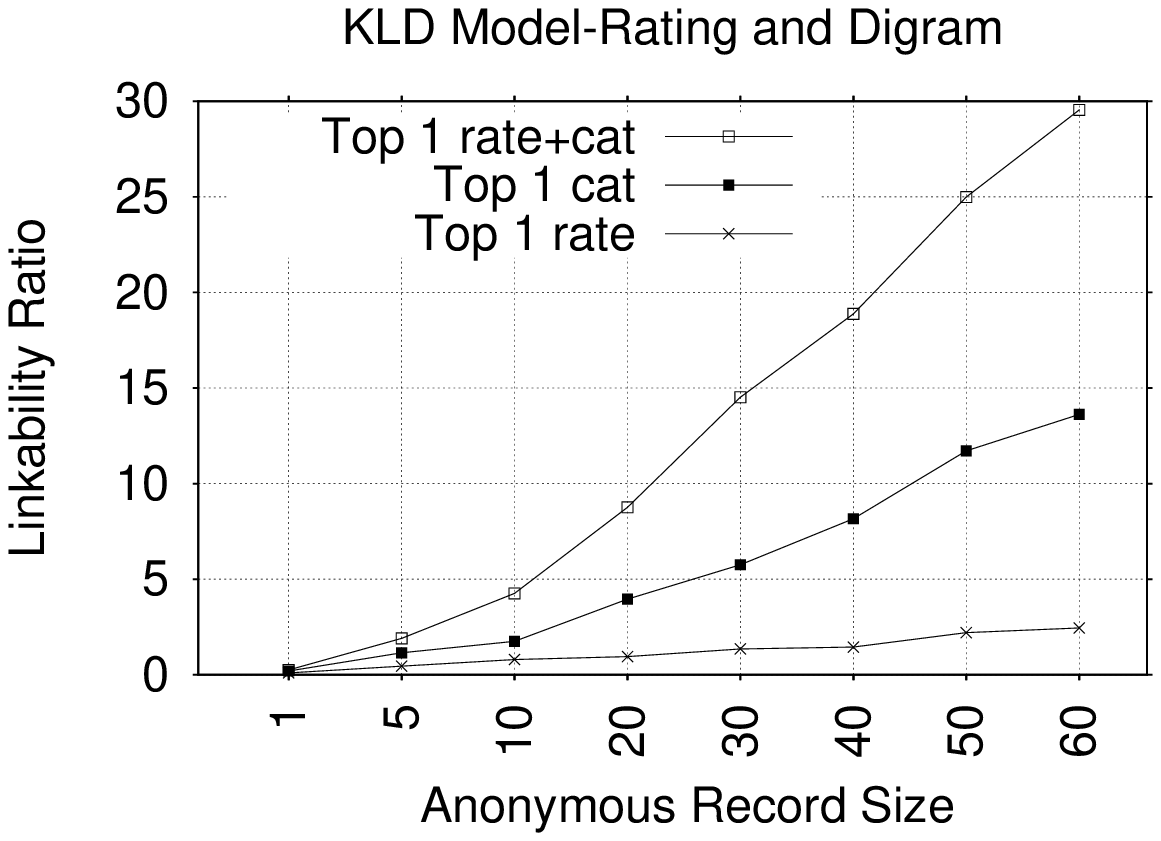}}
   \subfigure[]{\label{fig:r-c-kl}\includegraphics[scale=0.3]{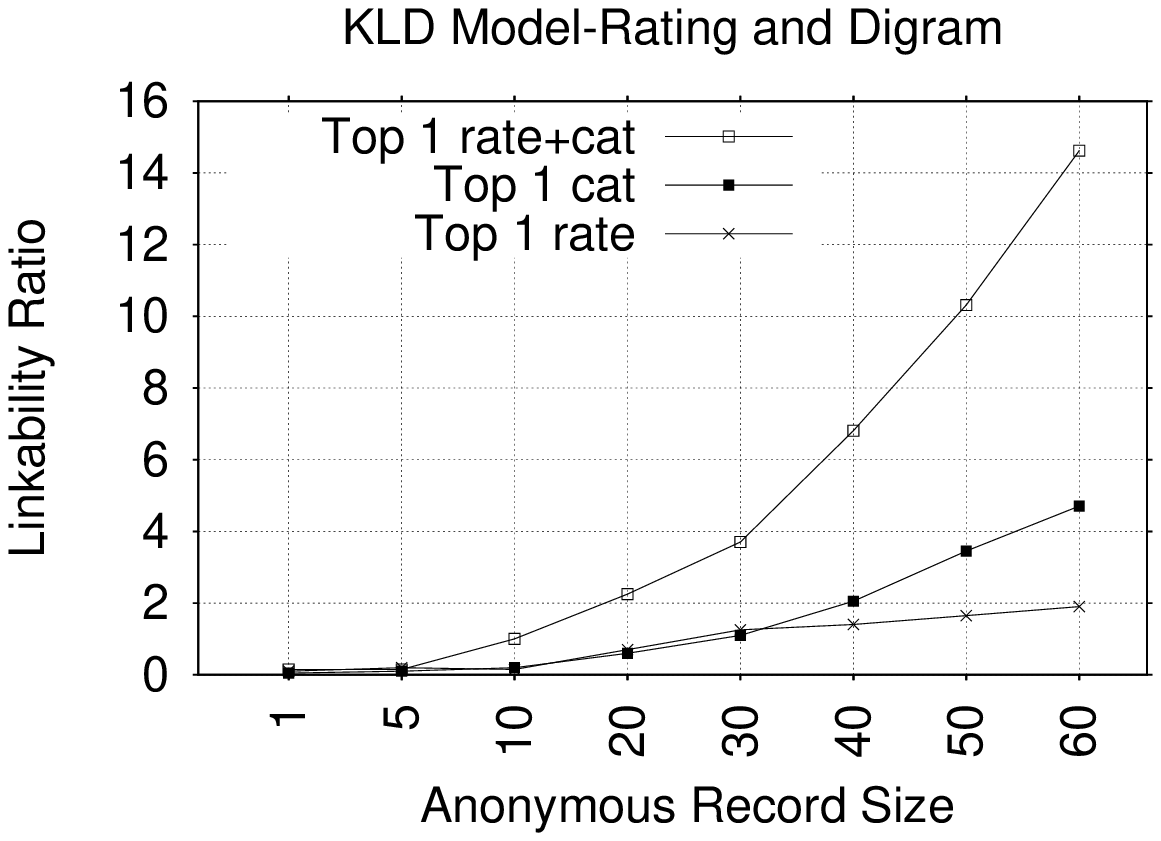}}
  \caption{LRs of NB and KLD  for combining ratings and categories}
  \label{fig:r-c}
\end{figure}

Figures \ref{fig:r-c-nb} and \ref{fig:r-c-kl} show Top-1 plots for NB and KLD models when rating and 
category tokens are combined or used separately. Clearly, combining the tokens significantly improves 
LRs in several record sizes. In NB, the gain in Top-1 LRs ranges from 2.5-15.9\%/3.5-27.1\% 
over the category/rating based model for most record sizes. For example, LRs increase from 5.8(1.4)\% 
and 13.6(2.5)\%  in category(rating) based model to 14.5\% and 29.5\% in NB combined model for 
record sizes of 30 and 60, respectively. In Top-50, the LR could reach as high as 87.7\%, 
versus 68.2(34.8)\% in the category (rating) based model for a record size of 60.   

In KLD, the gain in Top-1 LRs ranges from 1.7-9.9\%/1.6-12.7\% over category/rating based model for 
most record sizes. For example, it leaps from 1.1(1.3)\% and 4.7(1.9)\% in category (rating) based model  
to 3.7\% and 14.6\%  in KLD combined model for record sizes of 30 and 60, respectively. The gain is even 
higher in Top-50 where it reaches 69.1\%, versus 38.3(33.3)\% in the category (rating) based model 
for a record size of 60. These results show that combining rating and category tokens is very 
effective in increasing LRs in both NB and KLD models. 

\subsubsection{Combining Lexical with Non-Lexical Tokens}
\label{sec-results-uni-di-rate-cate-comb}
\begin{figure}[t]
  \centering
  \subfigure[]{\label{fig:ng-1-r-c-nb}\includegraphics[scale=0.3]{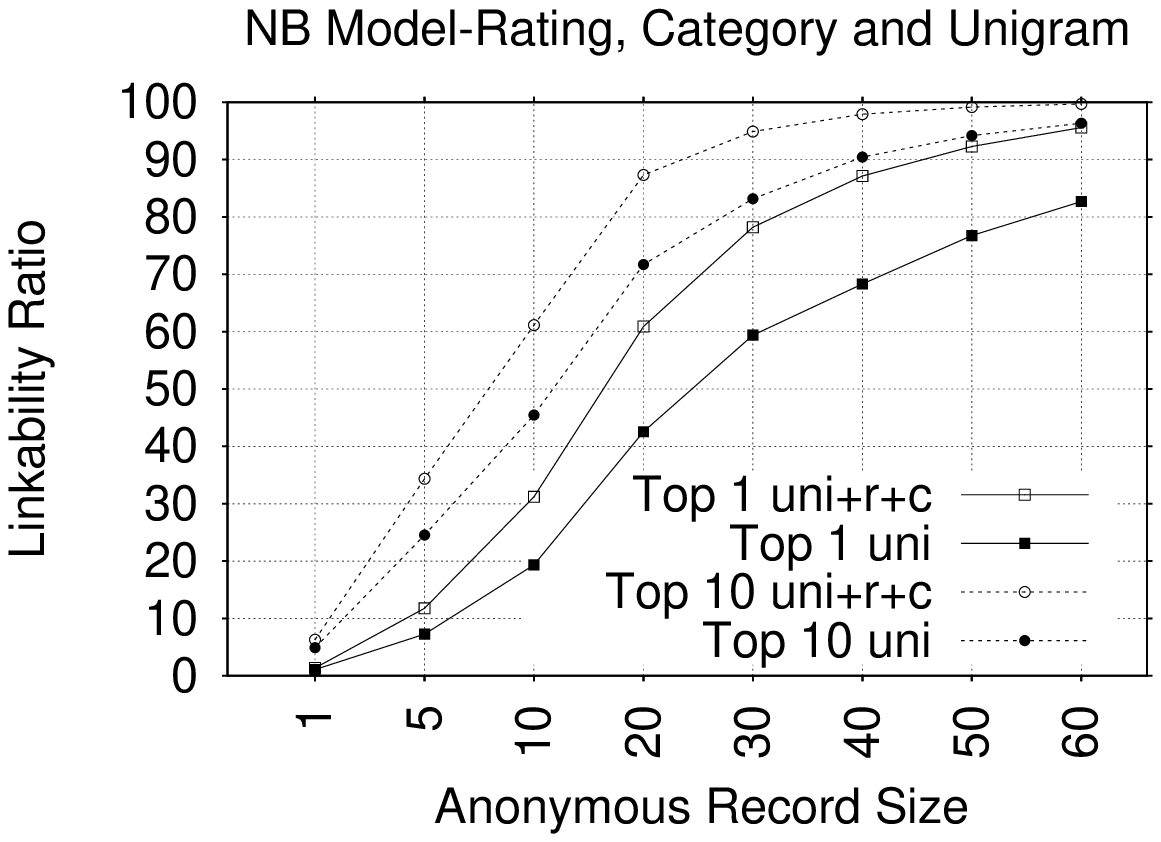}}
   \subfigure[]{\label{fig:ng-1-r-c-kl}\includegraphics[scale=0.3]{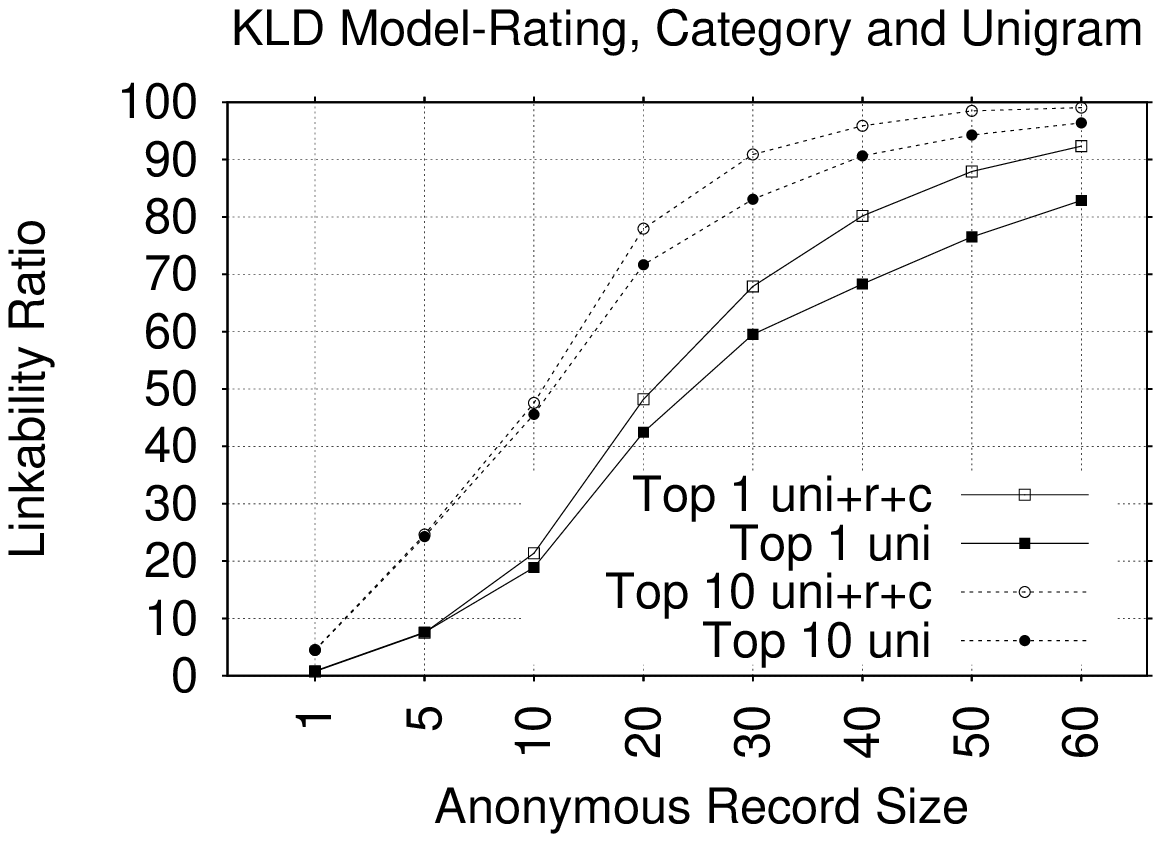}}
  \subfigure[]{\label{fig:ng-2-r-c-nb}\includegraphics[scale=0.3]{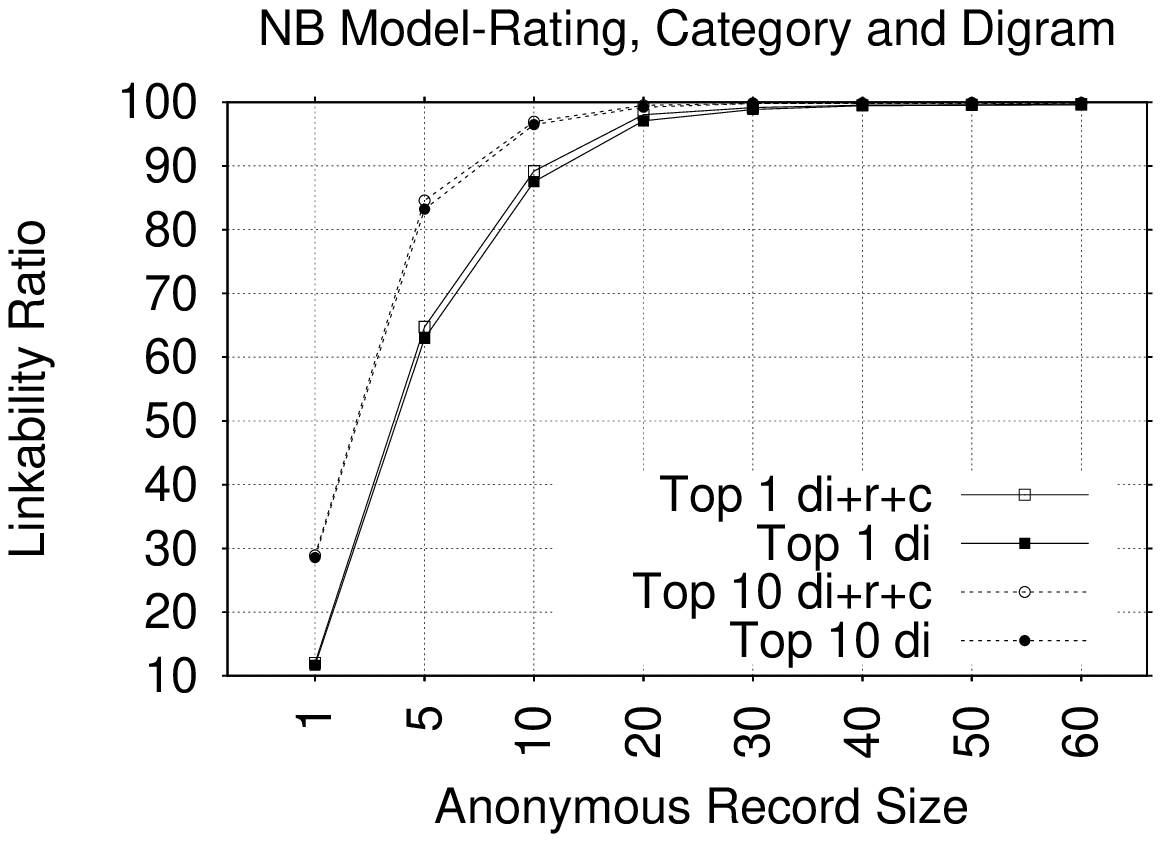}}
  \subfigure[]{\label{fig:ng-2-r-c-kl}\includegraphics[scale=0.3]{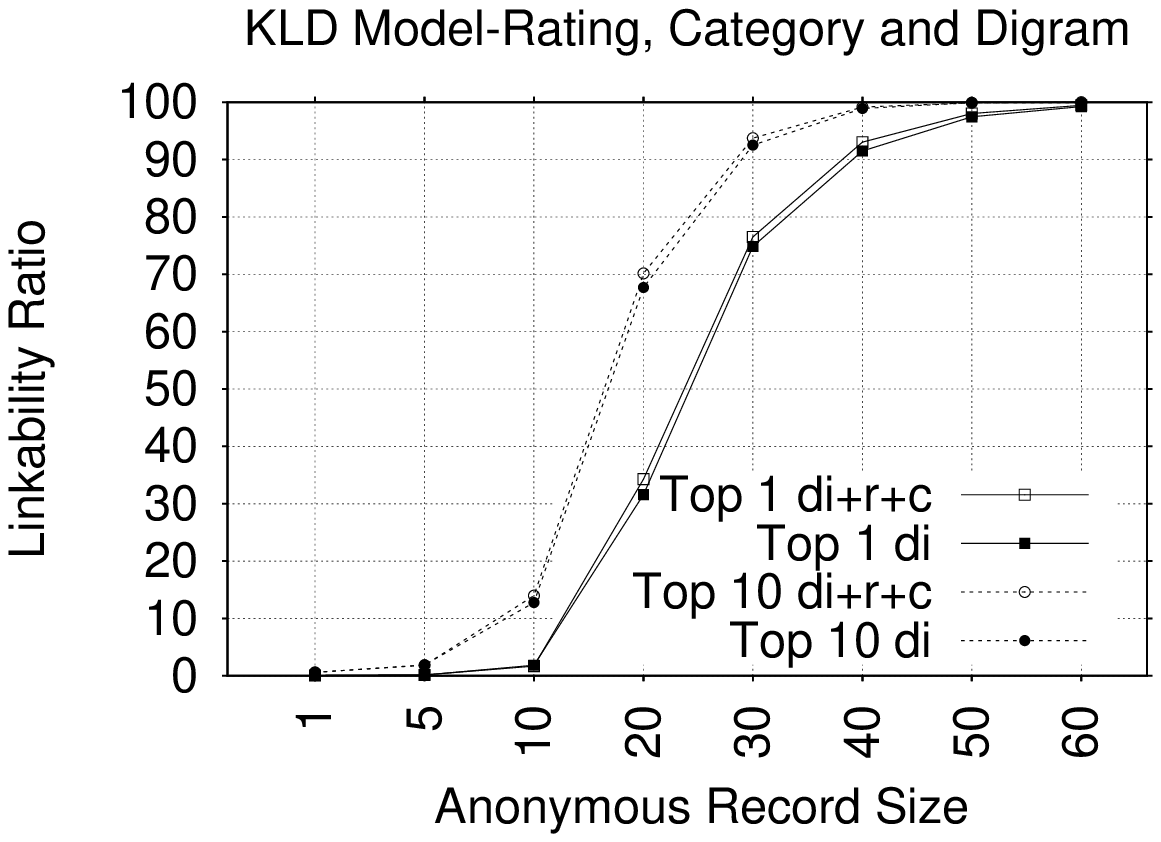}}
  \caption{LRs for NB and KLD for combining ratings and categories with unigrams or digrams}
  \label{fig:ng-2-r-c}
\end{figure}

Figures \ref{fig:ng-1-r-c-nb} and \ref{fig:ng-1-r-c-kl} show Top-1 and Top-10 plots in NB and KLD models of 
unigram tokens before and after combining them with rating and category tokens. Adding non-lexical tokens 
to unigrams substantially increases LRs in several record sizes. In NB, the gain in Top-1 LRs ranges from 
0.25-18.9\% (1.4 - 15.7\% for Top-10 LRs). 
In KLD, the gain in Top-1 LRs ranges from 2.5-11.9\ (2-7.8\% in Top-10 LRs) for most record sizes. 
These findings shows how effective is combining the non-lexical tokens with the unigrams. 
In fact, we can accurately identify almost all ARs. 

Figures \ref{fig:ng-2-r-c-nb} and \ref{fig:ng-2-r-c-kl} show the effect of adding ratings and categories to digrams. 
The overall effect is minuscule: in NB (KLD) model, the increase in Top-1 LRs ranges from 
0.3-1.8\% (0.2-2.7\%)  for most record sizes. The increase is very similar in Top-10 plots. 

\subsubsection{Restricting Identified Record Size}
\label{sec:train-min-max}
\begin{figure}[t]
  \centering
  \subfigure[]{\label{fig:min-max-ng-2-r-c-nb}\includegraphics[scale=0.3]{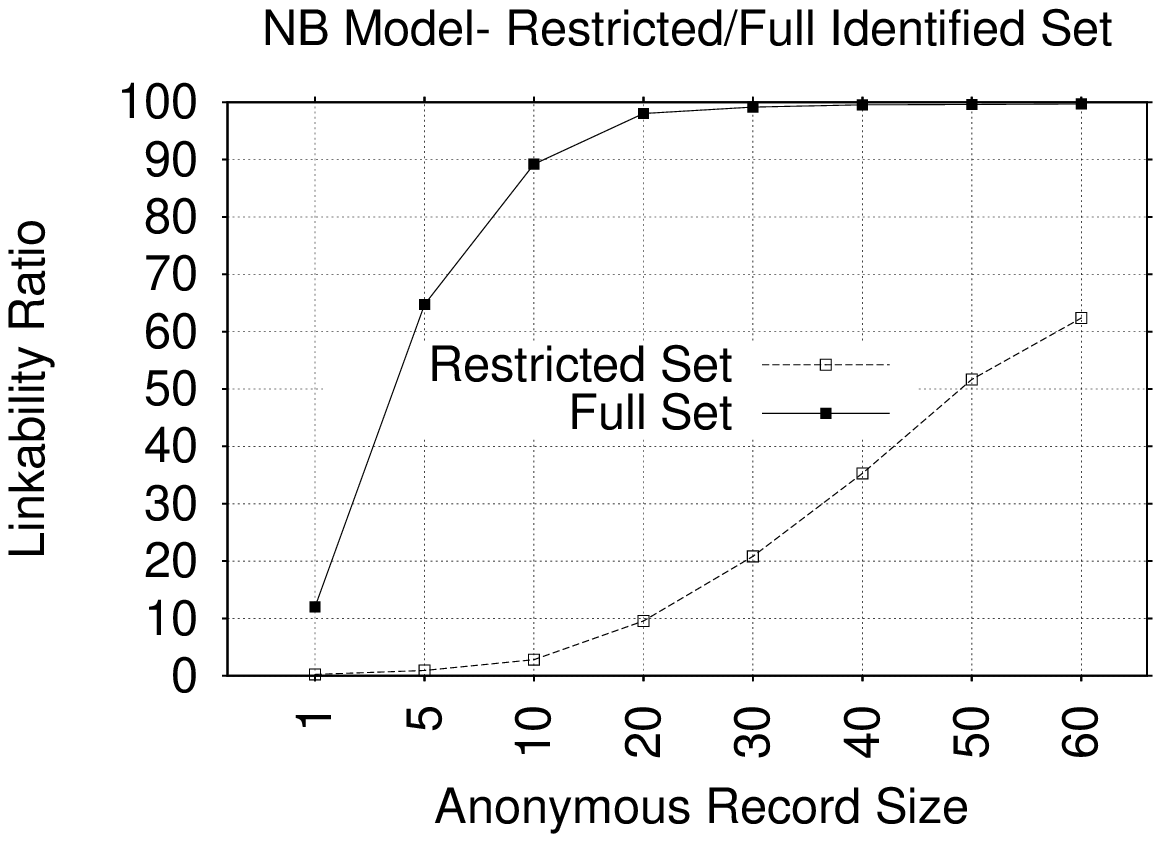}}
  \subfigure[]{\label{fig:min-max-ng-2-r-c-kl}\includegraphics[scale=0.3]{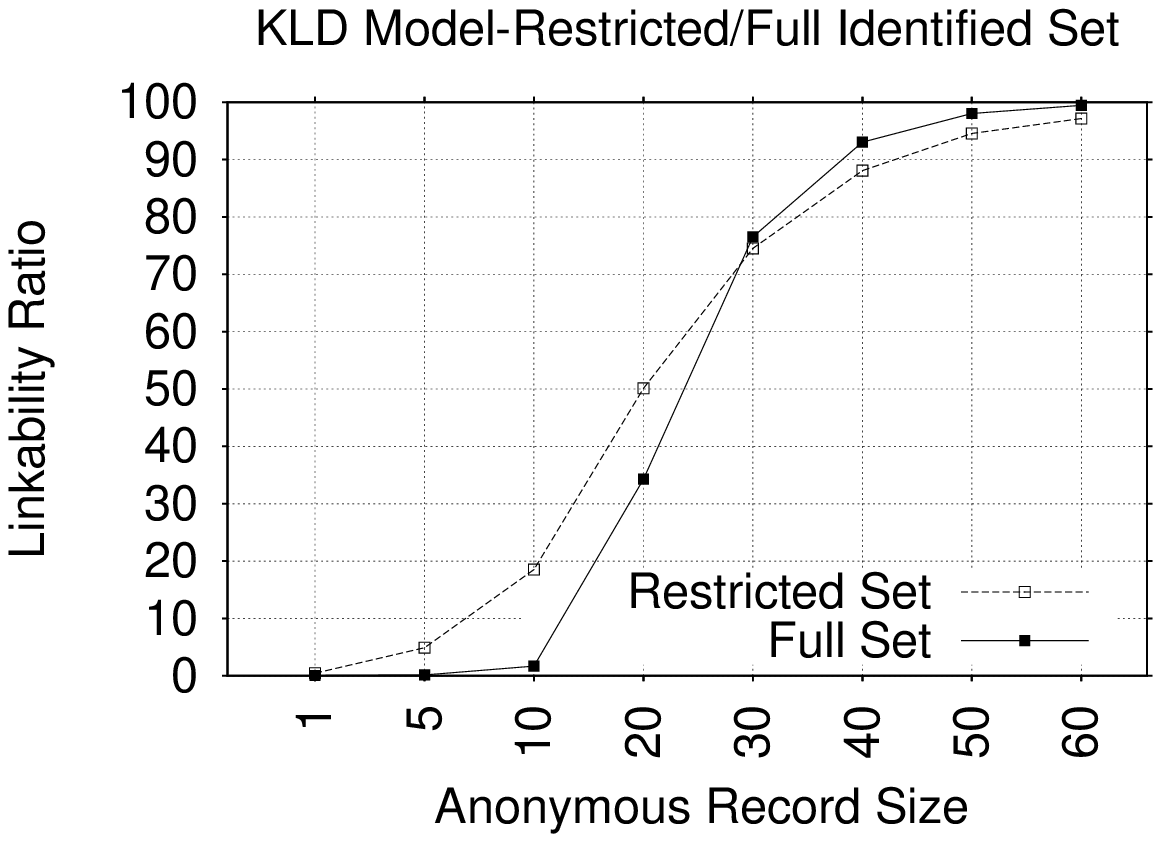}}
  \caption{LRs for NB and KLD in full and restricted identified set}
  \label{fig:min-max-ng-2-r-c}
\end{figure}

In previous sections, our analysis was based on using the full data set. That is, except for the 
anonymous part of the data set, we use all of the user reviews as part of our identified set. 
Although the LR is high in many cases, it is not clear how the models will perform when we 
restrict the IR size. To this end, we re-evaluate the models with the same problem settings, however,
with a restricted IR size. We restrict the IR size to the AR size; both randomly selected without replacement. 

Figures \ref{fig:min-max-ng-2-r-c-nb} and \ref{fig:min-max-ng-2-r-c-kl} show two Top-1 plots in NB and KLD 
models: one plot corresponds to the restricted identified set and the other -- to the full set. Tokens used in the models consist of digrams, ratings and categories (since this combination gives the highest LR). Unlike the previous sections, where NB and KLD behaved similarly, the two models now behave differently when 
restricting the identified set. While NB performs better than KLD on the full set, the latter performs much better  
than NB when the identified set is restricted. In fact, in some cases, KLD performs better when the set is 
restricted. 

%
The reason for this improved KLD performance might be the following: in the symmetric KLD distance function, the 
distributions of both the IR and AR have to be very close in order to match regardless of the size of the IR; unlike 
the NB, where larger training sets would lead to better estimates of the token probabilities and thus more accurate 
predictions.    

In KLD, we achieve high LRs for many record sizes. For example, Top-1 LRs in the restricted set are 74.5\%, 88\% 
and 97.1\% when the anonymous (and identified) record sizes are 30, 40 and 60, respectively. Whereas, 
 the LRs in the full set for the same AR sizes are: 76.5\% , 93\% and 99.4\%. When the record size is less than 30, 
KLD performs better in the restricted set than the full one. For example, when the AR size is 20, the LR in the 
restricted set is 50.1\% and 34.3\% in the full set. In NB, Top-1 LR in the restricted set is lower than the full set. 
For instance, it is 20.8\%, 35.3\% and 62.4\% for AR sizes of: 30, 40 and 60, respectively. Whereas, for the 
same sizes, the LR is more than 99\% in the full set.         

This result has one very important implication: even with very small IR sizes, many anonymous users 
can be identified. For example, with only IR and AR sizes of only 30, most users can be accurately linked 
(75\% in Top-1 and 90\% in Top-10). This situation is very common since many real-world users
generate 30 or more reviews over multiple sites. Therefore, even reviews from less prolific accounts 
can be accurately linked.   

\subsubsection{Improvement II: Matching all ARs at Once}
\label{sec:inter-model}
We now experiment with another natural strategy of attempting to match all ARs at once.

\subsubsection{Methodology}
\label{sec-method-once}
%
\begin{figure}[t]
\footnotesize
\begin{center}
\begin{tabular}{rl}
\hline\hline
\multicolumn{2}{l}{\textbf{Algorithm $Match\_All$:} Pseudo Code}\\
\hline
\textbf{Input}:& (1) Set of ARs: $S_{AR}=\{AR_1, AR_2, ...,AR_n\}$ \\
               & (2) Set of reviewer-ids / identified records:\\
	       & $S_{IR}=\{IR_1, IR_2, ..., IR_n\}$ \\
                &(3) Set of matching lists for each AR: \\
                 & $S_L=\{List_{AR_1},..,List_{AR_n}\}$ \\
\textbf{Output}:& Matching list: $S_M=\{(IR_{i_1},AR_{j_1}),É(IR_{i_n},AR_{j_n})\}$\\
1: & set $S_M=\emptyset$\\
2: & While $|S_{AR}| \neq 0$: \\
3: & \hspace{5 mm} Find $AR_i$ with smallest $SymD_{KLD}$ in all lists in $S_L$\\
4: & \hspace{5 mm} Get corresponding reviewer-id $IR_j$\\
5: & \hspace{5 mm} Add $(IR_j, AR_i)$ to $S_M$\\
6: & \hspace{5 mm} Delete $AR_i$ from $S_{AR}$\\
7: & \hspace{5 mm} Delete $List_{AR_i}$ from $S_L$\\
8: & \hspace{5 mm} For each $List_t$ in $S_L$, \\
9: & \hspace{10 mm} Delete tuple containing $IR_j$ from $List_t$\\
10:& \hspace{5 mm} End For\\
11:& End While\\
\hline
\end{tabular}
\end{center}
NOTE 1: $List_{AR_i}$ in $S_L$ is a list of pairs $(IR_j, V_{ij})$ where $V_{ij}=SymD_{KLD}(IR_j, AR_i)$, for all $j$

NOTE 2: $List_{AR_i}$ is sorted in increasing order of $V_{ij}$, i.e., $IR_j$ with lowest $SymD_{KLD}(IR_j, AR_i)$
at the top.
\caption{\label{fig:inter-kl-alg} Pseudo-Code for matching all ARs at once.}
\end{figure}
In the previous section, we focused on linking one AR at a time. That is,
ARs were independently and incrementally linked to IRs (accounts/reviewer-ids).
One natural direction for potential improvements is to attempt to link all ARs at the same time.
To this end, we construct algorithm $Match\_All()$ in Figure \ref{fig:inter-kl-alg} as an add-on to
the KLD models suggested in previous sections.

$SymD_{KLD}(IR_j, AR_i)$ symmetrically measures the distance between their 
($IR_j$'s and $AR_i$'s) distributions. Since every $AR$ maps to a distinct $IR$ ($AR_i$ maps to $IR_i$), 
it would seem that lower $SymD_{KLD}$ would lead to a better match. 
We use this intuition to design $Match\_All()$.
As shown in the figure, $Match\_All()$ picks the smallest $SymD_{KLD}(IR_j, AR_i)$ as the map 
between $IR_j$ and $AR_i$ and then deletes the pair $(IR_j, V_{kj})$ from all remaining lists in 
$S_L$. The process continues until we compute all matches. Note that, for any $List_{AR_k}$, 
$(IR_j, V_{kj})$ is deleted from the list only when there is another pair $(IR_j, V_{lj})$ in
$List_{AR_l}$, such that $SymD_{KLD}(IR_j, AR_l) \le SymD_{KLD}(IR_j, AR_k)$, and $IR_j$ 
has been selected as the match for $AR_l$.The output of the algorithm is a match-list: 
$S_M=\{(IR_{i_1},AR_{j_1}),É(IR_{i_n},AR_{j_n})\}$.

We now consider how $Match\_All()$ could improve the LR. Suppose that we have two ARs: 
$AR_i$ and $AR_j$
along with corresponding sorted lists $L_i$ and $L_j$ and assume that $IR_i$ is at the top of each list.
Using only KLD, we would return $IR_i$ for both ARs and thus miss one of the two.
Whereas, $Match\_All$, would assign $IR_i$ to {\bf only} one AR -- the one with
the smaller $SymD_{KLD}(IR_i, ...)$ value. We would intuitively suspect that 
$SymD_{KLD}(IR_i, AR_i) < SymD_{KLD}(IR_i, AR_j)$ since $IR_i$ is the right match for 
$AR_i$ and thus their distributions would probably be very close. 
If this is the case, $Match\_All$ would delete $IR_i$ (erroneous match) from the top of 
$L_j$ which could help clearing up the way for $IR_j$ (correct match) to the top of $L_j$. 

We note that there is no guarantee that $Match\_All()$ will always work: one mistake in early rounds 
would lead to others in later rounds. We believe that $Match\_All()$ works better if 
$SymD_{KLD}(IR_i, AR_i) < SymD_{KLD}(IR_j, AR_i)$ ($j\neq~i$) holds most of the time.

In the next section, we show the results of $Match\_All()$ when we experiment with the KLD model with 
digram, rating and category tokens.\footnote{We also tried $Match\_All()$ with the NB model and it did 
not improve the LR.}.

\subsubsection{Results}
\label{sec-results-once}
\begin{figure}[t]
  \centering
  \subfigure[]{\label{fig:inter-kl-min}\includegraphics[scale=0.3]{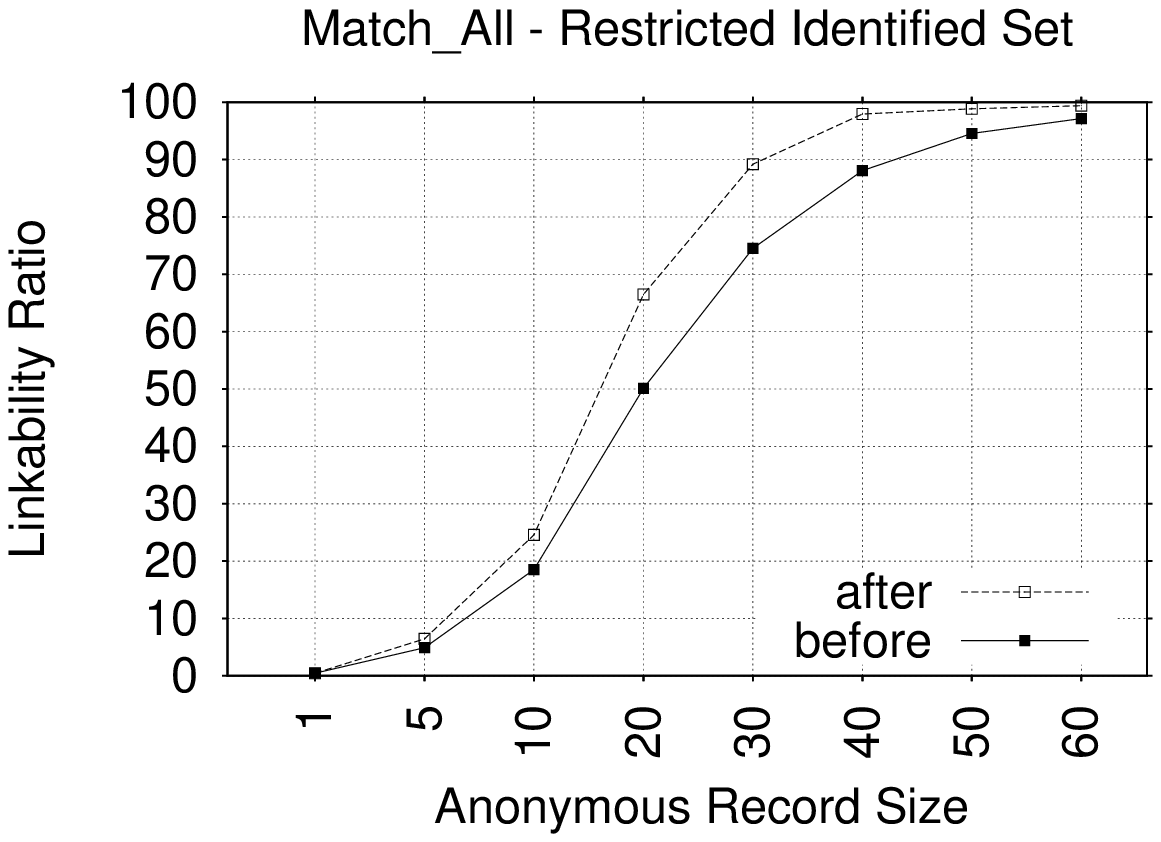}}
  \subfigure[]{\label{fig:inter-kl-max}\includegraphics[scale=0.3]{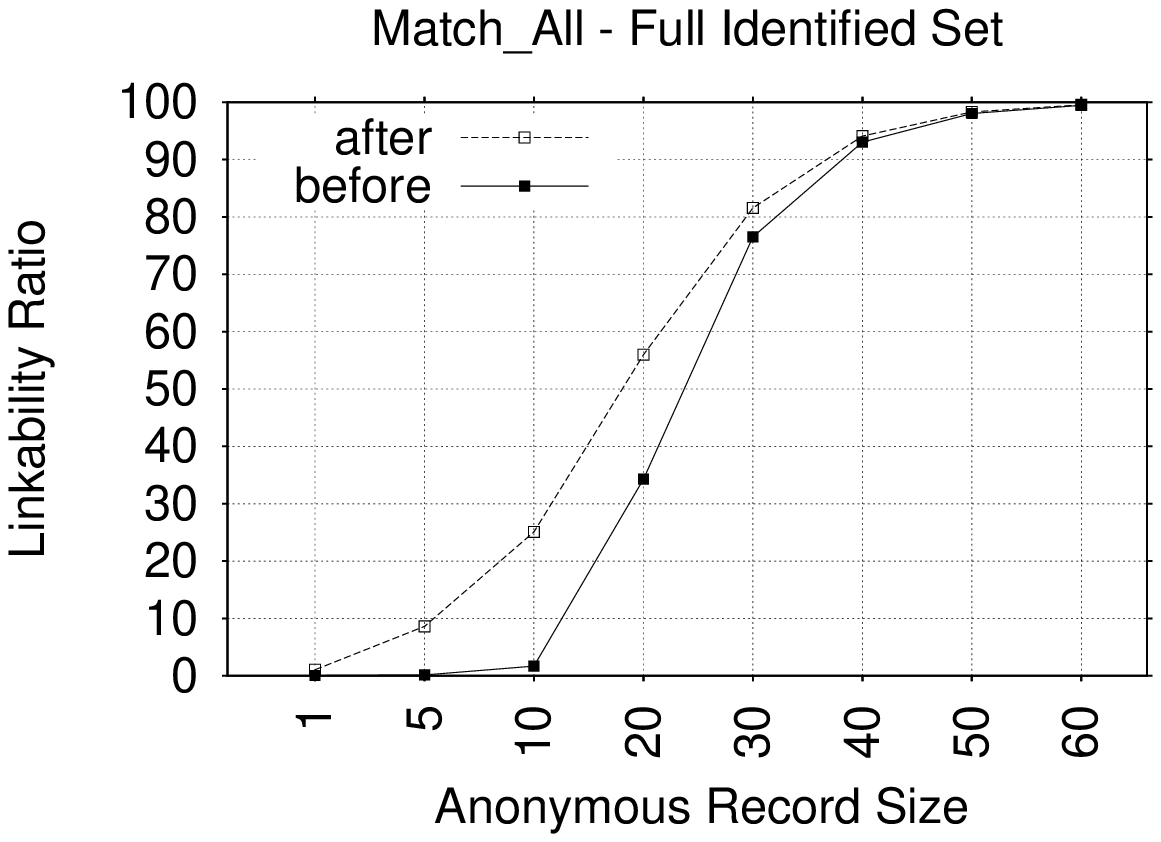}}
  \caption{Effects of $Match\_All()$  on LRs in full and restricted identified set: before and after plots}
  \label{fig:inter-kl-model}
\end{figure}
Figures \ref{fig:inter-kl-min} and \ref{fig:inter-kl-max} show the effect of $Match\_All()$ on Top-1 LRs in both the  
restricted identified set and the full identified set, respectively. The combination of diagram, rating and category 
tokens are used.  Each figure shows two Top-1 plots: one for the LR after using $Match\_All$ and the other  --
for the LR before using it. Clearly, $Match\_All$ is effective in improving the LR for almost all record sizes. For the 
restricted set, the gain in the LR ranges from 1.6-16.4\% for nearly all AR sizes. 
A Similar increase is observed in the full set that ranges from 1-23.4\% for most record sizes. 
This shows that the $Match\_All$ is very effective when used with diagram, rating and category tokens. 
The privacy implication of $Match\_All$ is important as it significantly increases the LR for small ARs in the 
restricted set. This shows that privacy of less prolific users is exposed even more with $Match\_All$.

\subsection{Study Summary}
\label{sec:summary-analysis}
We now summarize the main findings and conclusions of our study.
\begin{enumerate}
\item The LR becomes very high -- reaching up to $\sim$ 99.5\% in both KLD and NB when using only 
digram tokens. (See Section \ref{sec:uni-di-results}).   
\item Surprisingly, using only unigrams, we can link up to 83\% in both NB and KLD models, 
with 96\% in Top-10. (See Section \ref{sec:uni-di-results}). This suggests that reviewers 
expose a great deal merely from their single letter distributions.
\item Even with small record sizes, we accurately link a significant ratio of ARs. Specifically, for 
AR sizes of 5 and 10 (using NB with diagrams), we can accurately link 63\% and 88\% ARs, 
respectively. (See Section \ref{sec:uni-di-results}).
\item Rating and category tokens are more useful if combined where 88\%/69\% of ARs (size 60) 
fall into Top-50 in NB/KLD. (See Section \ref{sec-results-rate-cate-comb}).
\item Non-lexical tokens are very useful in tandem with lexical tokens, especially, the unigram: 
we observe a $\sim$19\%/12\% Top-1 LR increase in NB/KLD for some cases.
(See Section \ref{sec-results-uni-di-rate-cate-comb}).
\item Relying only on unigram, rating and category tokens, we can accurately link 96\%/92\% of the ARs 
(size 60) in NB/KLD. (See Section \ref{sec-results-uni-di-rate-cate-comb}).
\item Restricting the IR size does not always degrade linkability. In KLD, we can link as many as 97\% 
ARs when the IR size is small. (See Section \ref{sec:train-min-max}). 
\item  Linking all ARs at once (instead of each independently) helps improve accuracy.
The gain is up to 16/23\% in restricted/full set. (See Section \ref{sec-results-once}).
\item Generally, NB performs better than KLD when we use the full identified set and KLD 
performs better when we use the restricted identified set. 
\end{enumerate}

section{Discussion}
\label{sec:disc}
\textbf{Implications:} We believe that the results of, and techniques used in, this study have several 
implications. First, we demonstrated the practicality of cross-referencing accounts (and reviews) among
multiple review sites. If a person contributes to two sites under two identities, it is highly 
likely that sets of reviews from these sites can be linked.
This could be quite detrimental to contributors' privacy.  

The second implication is the ability to correlate -- on the same review site -- multiple accounts 
that are in fact manipulated by the same person. 
This  could make our techniques very useful in detecting review spam \cite{opinion-spam-analysis},
whereby a contributor authors reviews under different accounts to tout (also self-promote) or criticize a
product or a service. 

\textbf{Prolific Users:} While there are clearly many more occasional (non-prolific) reviewers than prolific ones, 
we believe that our study of prolific reviewers is important, for two reasons. First, the number of 
prolific contributors is still quite large. For example, from only one review site -- Yelp -- we identified 
$\sim$ $ 2,000$ such reviewers. Second, given the spike of popularity of review sites \cite{yelpstat}, we believe
that, in the near future, the number of such prolific contributors will grow substantially.  
Also, even many occasional reviewers, with the passage of time, will enter the ranks of 
``prolific'' ones, i.e., by slowly accumulating a sufficient corpus of
reviews over the years. Nevertheless, our study suggests that privacy is not high even for 
non-prolific users, as discussed in Section \ref{sec:inter-model}. For example, when both IR and AR 
sizes are only 20 (i.e., total per user contribution is 40 reviews), we can accurately link half of 
anonymous records to their reviewers.

\textbf{Anonymous Record Size:} Our models perform best when the AR size is 60. However, for every 
reviewer in our dataset,  60 represents less than 20\% of that person's total number of reviews. 
Also, using NB coupled with digram, rating and category tokens, we can accurately link most 
anonymous records when the AR size is only 10. Interestingly, the AR size of 10 
represents only 3\% of the minimum user contribution.   

\textbf{Unigram Tokens:} While our best-performing models are based on digram tokens, we also
obtain high linkability results from unigram tokens that reach up to 83\% (96\% in the Top 10) in NB or KLD. 
The results improve to 96/92\% when we combine unigrams with rating and category tokens. Note that 
the number of tokens in unigram-based models is 59 (26) tokens with (without) combining them 
with rating and category tokens. Whereas, the number of tokens in diagram-based models is 676 (709 
when combined with rating and category tokens). This makes linkability accuracy
based on unigram models very comparable to its diagram counterpart, while the number of tokens is 
significantly fewer. 
%
This implies a substantial reduction in resources and processing power in unigram-based
models which would make them scale better. For example, if we assume that the attacker wants to link
a set of anonymous reviews to {\em many} large review datasets, unigram-based models would 
scale better, while maintaining the same level of accuracy.    

\textbf{Potential Countermeasures:}
One concrete application of our techniques is via integration with the review site's
front-end software in order to provide feedback to authors indicating the degree of linkability
of their reviews. For example, when the reviewer logs in, a linkability nominal/categorical value 
(e.g. high, medium, and low) could be shown indicating how his/her last $k$ reviews (where $k$ is small,
e.g., 5 or 10) are linkable to the rest. It would then be up to to the individual to maintain or modify their 
reviewing patterns to be less linkable. Another way of countering linkability is for the front-end
software to automatically suggest a different choice of words that are less revealing (less personal) 
and more common among many users. We suspect that, with the use of such words, reviews would be less 
linkable and lexical distributions for different users would be more similar. 

%

\section{Future Work}
\label{sec:futurework}
Although our results point at high linkability of reviews, there remain many open
questions. First, the anonymous records are not highly linkable when their sizes are very small, e.g., 1 or 5. 
As part of future work, we plan to improve linkability on very small anonymous records. In addition, 
although we take advantage of ratings and categories to boost LRs, we need to further explore usage of 
other non-textual features, such as sub-categories of places, products and services reviewed as well as
the length of reviews. In fact, it would be interesting to see how the LR can be improved
without resorting to lexical features, since they generally entail heavy processing. We also plan to 
implement countermeasures techniques described in Section \ref{sec:disc} and examine their 
efficacy.   

Moreover, we plan to investigate LRs in other preference databases, such as music/song ratings, 
and check whether contributors inadvertently link their reviews through preferences. 
It would be interesting to see how to leverage techniques used in recommender systems 
(for future rating prediction) to increase LRs.  

In Section \ref{sec:inter-model}, we showed how to improve LRs by linking all anonymous records at once. 
We intend to further investigate the effect on the LR of the number anonymous records 
when each record belongs to a different reviewer.

\section{Related Work}
\label{sec:related}
Many authorship analysis studies have appeared in the literature. Among the most prominent
recent studies are: \cite{unified-data-mine,  author-ident-framework-rong, writeprints-abbasi}. 
The study in \cite{unified-data-mine} proposed
techniques that extract frequent pattern write-prints that characterized one (or a group of)
authors. The best achieved accuracy was 88\% when identifying an author, from a single anonymous message,
from a small set of four and with training set size of forty messages per author. In 
\cite{author-ident-framework-rong}, a framework for author identification for on-line messages was 
introduced where four types of features are extracted: lexical, syntactic, structural and content-specific.
Three types of classifiers were used for author identification: 
Decision Trees, Back Propagation Neural Networks and Support Vector machines. 
The last one outperformed the others, achieving 97\% in a set of authors less than 20. The work in 
\cite{writeprints-abbasi} also considered author identification and similarity detection by incorporating 
a rich set of stylistic features along with a novel technique(based on Karhunen-Loeve-transforms) to 
extract write-prints. These techniques performed well, reaching as high as 91\% in 
identifying the author of anonymous text from a set of 100. The
same approach was tested on a large set of Buyer/Seller Ebay feedback comments collected from Ebay.
Such comments typically reflect one's experience when dealing with a buyer or a seller. Unlike our general-purpose
reviews, these comments do not review products, services or places of different categories. Additionally,
the scale of the problem was different and the analysis was performed for 100 authors, whereas,  our analysis
involved $\sim~2,000$ reviewers.

A problem very similar to ours was explored in \cite{herbert-deanonymizer}. It focused on identifying authors 
based on reviews in both single- and double-blinded peer-reviewing processes. Na\"ive Bayes classifier was
used -- along with unigrams, bigrams and trigrams -- to identify authors and the best result was  around 90\%. 
In \cite{blind-review}, citations of a given paper were used to identify its authors. The data set was a very 
large archive of physics research papers (KDDCUP 2003 physics-paper archive). Authors were identified 
40-45\% of the times.  In \cite{ngram-bayes-topic-fixed}, authorship analysis was performed on a set of 
candidate authors who wrote on
the same topics. Specifically, analysis was done on movie reviews of five reviewers on the same five movies.
Although reviews similar to ours were used, there were significant differences. We use over $1,000,000$ reviews
by $\sim~2,000$ authors, whereas, only 25 reviews by 5 authors were used in \cite{ngram-bayes-topic-fixed}. A 
related result \cite{gender-movie} studied the problem of inferring the gender of a movie reviewer from
his/her review. Using logistic regression \cite{log-regress} along with features derived from the writing style,
content, and meta-data of the review, accuracy of up to 73.7\% was achieved in determining the correct gender.
The goal of this study was clearly quite different from ours. For a comprehensive overview of
authorship analysis studies, we refer to \cite{author-survey}.

While all aforementioned results are somewhat similar to our present work, there are some notable 
differences. First, we perform authorship identification analysis in a context that has not been extensively 
explored -- user reviews. User reviews are generally written differently
from other types of writing, such as email and research papers. In a review, the author generally 
assesses something  and thus the text conveys some evaluation and personal opinions. A review usually 
conveys information about personal taste, since most people tend to review things of interest to them. In 
addition, reviews contain other non-textual information, such as the ratings and categories of things being 
reviewed. These types of extra information provide added leverage; recall that, as discussed earlier,  
ratings and categories are particularly helpful in increasing the overall linkability ratio. Second, our 
problem formulation is different. We study linkability of reviews (and user-ids of their authors) in the presence 
of a large number of prolific contributors where the number of anonymous reviews could be more than one 
(up to 60 reviews). Whereas, most prior work attempts to identify authors from a small set of authors, each 
with small sets of texts where the number of anonymous documents/messages is one.       

Some work has been done in recovering authors based on their ratings, using external knowledge. In particular,
\cite{you-what-u-say} studied author linkability with two different databases; one public and the other -- private.
Several techniques were used to link authors in public forums (public) who state their opinions and rating
about movies to reviewers who contribute to a sparse database (private) of movie ratings.  A related result
\cite{deanonymize-netflix} considered
anonymity in high-dimensional and sparse data sets of anonymized users. First, it presented a general definition and model
of privacy breaches in such sets.Second, a statistical de-anonymization attack
was presented that was resilient to perturbation. Third, this attack was used to de-anonymize the Netflix \cite{netflix}
data set. Note that the problem formulation of these two results differs from ours.
They studied anonymity in the presence of an external source of public information. Whereas,
our work does not rely on any external sources.

Last but not least, other related research effort assessed authenticity of reviews \cite{opinion-spam-analysis}. 
It explored the problem of identifying spam. Results demonstrated that spam reviews were 
prevalent and a counter-measure based on logistic regression was proposed.

\section{Conclusion}
\label{sec:conclusion}
Large numbers of Internet users are becoming frequent visitors and contributors to various review sites. 
At the same time, they  are concerned about their privacy. In this paper, we study linkability of reviews. Based on a large set of 
reviews, we show that a high percentage (99\% in some cases) are linkable, even though we use very simple models and very 
simple features set. Our study suggests that users reliably expose their identities in reviews. This has certain important 
implications for cross-referencing accounts among different review sites and detecting people who write reviews 
under different identities.  Additionally,  techniques used in this study could be adopted by review sites to give 
contributors feedback about linkability of their reviews.

\eject

\bibliographystyle{abbrv}
\bibliography{paper}

\end{document}